\documentclass[10pt,letterpaper]{article}
\usepackage[top=0.85in,left=2.75in,footskip=0.75in,marginparwidth=2in]{geometry}

\usepackage[utf8]{inputenc}

\usepackage{natbib}
\setcitestyle{authoryear}

\usepackage{nameref,hyperref}

\usepackage[right]{lineno}

\usepackage{microtype}
\DisableLigatures[f]{encoding = *, family = * }

\usepackage{url} 
\usepackage{lineno}

\usepackage{bm}
\usepackage{amsmath}
\usepackage{algorithm}
\usepackage{amsfonts}
\usepackage{amssymb}
\usepackage{multirow}

\raggedright
\usepackage{changepage}

\usepackage[aboveskip=1pt,labelfont=bf,labelsep=period,singlelinecheck=off]{caption}

\makeatletter
\renewcommand{\@biblabel}[1]{\quad#1.}
\makeatother

\usepackage{lastpage,fancyhdr,graphicx}
\usepackage{epstopdf}
\fancyheadoffset[L]{2.25in}
\fancyfootoffset[L]{2.25in}

\usepackage{color}

\definecolor{Gray}{gray}{.25}

\usepackage{graphicx}

\usepackage{sidecap}

\usepackage{wrapfig}
\usepackage[pscoord]{eso-pic}
\usepackage[fulladjust]{marginnote}
\reversemarginpar

\begin{document}
\vspace*{0.35in}

\begin{flushleft}
{\Large
\textbf{Real-Time Thermospheric Density Estimation Via Two-Line-Element Data Assimilation}
}
\newline
\\
\bf David J. Gondelach\textsuperscript{*} and Richard Linares
\\
\bigskip
\rm Department of Aeronautics and Astronautics, Massachusetts Institute of Technology, Cambridge, MA, USA
\\
\bigskip
* dgondela@mit.edu

\end{flushleft}

\subsection*{Abstract}
Inaccurate estimates of the thermospheric density are a major source of error in low Earth orbit prediction. To improve orbit prediction, real-time density estimation is required. In this work, we develop a reduced-order dynamic model for the thermospheric density by computing the main spatial modes of the atmosphere and deriving a linear model for the dynamics. The model is then used to estimate the density using two-line element (TLE) data by simultaneously estimating the reduced-order modes and the orbits and ballistic coefficients of several objects using an unscented Kalman filter. Accurate density estimation using the TLEs of 17 objects is demonstrated and validated against CHAMP and GRACE accelerometer-derived densities. Finally, the use of the model for density forecasting is shown.

\paragraph*{Keypoints}
\begin{itemize}
    \item Thermospheric density is estimated using a dynamic reduced-order thermosphere model and two-line element data.
    \item The two-line element data is assimilated using a Kalman filter to provide both density and uncertainty estimates.
    \item The estimated densities are validated against CHAMP and GRACE accelerometer-derived density data.
\end{itemize}

\section{Introduction}
Accurate knowledge of the thermospheric density is essential for orbit prediction in low Earth orbit and in particular for conjunction assessments. The most accurate models of the thermosphere are physics-based models, such as the Global Ionosphere-Thermosphere Model (GITM) \citep{RidleyEtAl2006} and the Thermosphere-Ionosphere-Electrodynamics General Circulation Model (TIE-GCM) \citep{qian2014ncar}. These models solve the continuity, momentum and energy equations for a number of neutral and charged components. Their modeling and prediction performance comes, however, at a high computational cost. The models are very high-dimensional, solving Navier-Stokes equations over a discretized spatial grid involving $10^4$-$10^6$ state variables and 12-20 inputs and internal parameters. In addition, to fully exploit the forecasting potential of physics-based models the schemes employed for data assimilation need to be improved \citep{sutton2018new}.

Empirical models \citep{jacchia1970new,hedin1987msis,PiconeEtAl2002,bowman2008new,bruinsma2015dtm}, on the other hand, capture the average behaviour of the atmosphere using low-order, parameterized mathematical formulations based on historical observations. A major advantage of empirical models is that they are fast to evaluate, making them ideal for drag and orbit computations. The accuracy of these empirical models is however limited \citep{he2018review}, especially during space weather events. Improved densities can be obtained by calibrating empirical density models using satellite data. The current Air Force standard is the High Accuracy Satellite Drag Model (HASDM) \citep{storz2005high}, which is an empirical model that is calibrated using observations of calibration satellites. These satellite observations are used to determine atmospheric model parameters based on their orbit determination solutions. Due to the lack of access to space surveillance observations, publicly available two-line element (TLE) data have been used in the past to estimate the thermospheric density \citep{picone2005thermospheric,emmert2006thermo} and calibrate empirical models \citep{cefola2004atmospheric,yurasov2005density,doornbos2008use,chen2019improved}. \citet{cefola2004atmospheric} and \citet{yurasov2005density} calibrated the GOST and NRLMSISE-00 density models using two scaling parameters based on the fitted ballistic coefficient (BC) values. \citet{doornbos2008use} estimated spherical harmonics coefficients to calibrate NRLMSISE-00 model using TLE-derived density estimates and \citet{sang2011modification,chen2019improved} adjusted the 187 coefficients of the DTM87 density model directly during orbit determination of multiple objects. On the other hand, \citet{crowley2017reducing} used TLE data for data assimilation in the physics-based Dragster model using Ensemble Kalman filtering. 
Except for the Dragster model, the calibrated empirical models have limited forecasting capability which reduces their effectiveness for orbit prediction.

Recently, a new methodology for modelling and estimating the thermosphere using reduced-order modeling was developed by \citet{mehta2017methodology} to overcome the high-dimensionality problem of physics-based models. The technique combines the predictive abilities of physics-based models with the computational speed of empirical models by developing a Reduced-Order Model (ROM) that represents the original high-dimensional system using a smaller number of parameters. 
The order-reduction is achieved using proper orthogonal decomposition (POD) \citep{golub1970singular,rowley2004model}, also known as principal component analysis (PCA) \citep{jolliffe2011principal}, empirical orthogonal functions (EOF) or Karhunen-Loeve expansion \citep{loeve1977elementary}.
The POD approach uses singular value decomposition (SVD) to compute spatial modes that are orthogonal with respect to each other. Using the POD modes, a ROM can be constructed from experimental or numerical data. In addition, POD modes have been used by \citet{matsuo2010principal} and \citet{sutton2012thermo} to study and model variations of the thermospheric density.
To model the dynamic thermosphere, a dynamic ROM was developed by \citet{mehta2018quasi} by determining the best fit linear dynamical system from density data using the recently developed Dynamic Mode Decomposition (DMD) technique \citep{schmid2010dynamic}. The DMD approach assumes a linear system ${\bm x}_{k+1} = A {\bm x}_k$, where ${\bm x}_k$ is the $k^\text{th}$ data sampled from a sequential dataset and $A$ is the unknown system matrix. In particular, Dynamic Mode Decomposition with control (DMDc) \citep{proctor2016dynamic} can include the effect of control and extends the DMD approach to systems with the form ${\bm x}_{k+1} = A {\bm x}_k+ B{\bm u}_k$ where ${\bm u}_k$ is the system input. The DMDc method was used by \citet{mehta2018quasi} to develop a quasi-physical dynamic ROM for the thermosphere. The application of the ROM approach for atmospheric density estimation was demonstrated by data assimilation of accelerometer derived mass density \citep{mehta2018new} and simulated GPS measurements \citep{mehta2018data} using Kalman filters. This technique enables both the accurate estimation of thermospheric density and forecasting of the future density through the ROM dynamic model. 

The benefits of this approach are that:
1) the modes contain most energy of the system and are orthogonal;
2) the modes can be estimated in real-time thanks to the dynamic model, which is not possible with a static model;
3) the dynamic model can be used for forecasting (for static models only the fitted coefficients can be extrapolated in time).

In this work, the reduced-order modelling technique for density estimation is further developed and TLE data is used to estimate the thermospheric density. The availability of TLE data for thousands of objects make them attractive for density estimation; however, the use of TLEs is challenging due to the limited accuracy of the orbital data. The density estimation using TLE data is achieved by simultaneously estimating the orbits and BCs of several objects and the reduced-order density state using an unscented Kalman filter.
The main contributions of the paper are:
\begin{enumerate}
\item Nonlinear space weather inputs are introduced to improve ROM prediction.
\item Two new ROM models based on the NRLMSISE-00 and JB2008 models are developed to extend the maximum altitude to 800 km.
\item Modified equinoctial elements are employed to express the orbit measurements.
\item Thermospheric densities are estimated using reduced-order modeling and TLE data.
\item Accurate density estimation over extended periods of time using a limited amount of TLE data through the use of the dynamic density models is demonstrated.
\item The estimated densities are validated against CHAMP and GRACE accelerometer-derived density data.
\end{enumerate}

The paper is structured as follows. First the development of a dynamic reduced-order density model described. After that, the estimation of the density via TLE data assimilation using an unscented Kalman is discussed. Then the performance of the ROM density prediction and estimation is assessed using simulated and real TLE cases, and finally conclusions are drawn.

\section{Methodology}
The neutral density estimation approach consists of two main components: 1) the development of a dynamic reduced-order model (ROM) for the thermosphere and 2) the calibration of the ROM through assimilation of TLE data.

\subsection{Reduced-order modeling}
The main idea of reduced-order modeling is to reduce the dimensionality of the state space while retaining maximum information. In our case, the full state space consists of the neutral mass density values on a dense uniform grid in latitude, local solar time and altitude. The goal is to develop a model for the density evolution over time. 
First, to make the problem tractable, the state space dimension is reduced using POD. Second, a linear dynamic model is derived by applying DMDc.

\subsubsection{Proper orthogonal decomposition} The concept of order reduction using POD is to project the high-dimensional system and its solution onto a set of low-dimensional basis functions or spatial modes, while capturing the dominant characteristics of the system. 
Consider the variation $\tilde{\mathbf{x}}$ of the neutral mass density $\mathbf{x}$ with respect to the mean value $\bar{\mathbf{x}}$:
\begin{linenomath*}\begin{equation}
    \tilde{\mathbf{x}}(\mathbf{s},t) = {\mathbf{x}}(\mathbf{s},t) - \bar{\mathbf{x}}(\mathbf{s})
    \label{eq:variationX}
\end{equation}\end{linenomath*}
where $\mathbf{s}$ is the spatial grid.
A significant fraction of the variance $\tilde{\mathbf{x}}$ can be captured by the first $r$ principal spatial modes:
\begin{linenomath*}\begin{equation}
    \tilde{\mathbf{x}}(\mathbf{s},t) \approx \sum_{i=1}^r c_i(t) \Phi_i(\mathbf{s})
\end{equation}\end{linenomath*}
where $\Phi_i$ are the spatial modes and $c_i$ are the corresponding time-dependent coefficients.
The spatial modes $\Phi$ are computed using a SVD of the snapshot matrix ${\bf X}$ that contains $\tilde{\mathbf{x}}$ for different times:
\begin{linenomath*}\begin{equation}
{\bf X}= \begin{bmatrix}
| & | & & | \\
\tilde{\mathbf{x}}_1 & \tilde{\mathbf{x}}_2 & \cdots & \tilde{\mathbf{x}}_m \\
| & | & & |
\end{bmatrix}
= \mathbf{U} \mathbf{\Sigma} \mathbf{V}^T
\label{eq:SVD}
\end{equation}\end{linenomath*}
The spatial modes $\Phi$ are given by the left singular vectors (the columns of $\mathbf{U}$).
The state reduction is achieved using a similarity transform:
\begin{linenomath*}\begin{equation}
    {\mathbf{z}} = \mathbf{U}_r^{-1} \tilde{\mathbf{x}}  = \mathbf{U}_r^T \tilde{\mathbf{x}} 
    \label{eq:orderReduction}
\end{equation}\end{linenomath*}
where $\mathbf{U}_r$ is a matrix with the first $r$ POD modes and ${\mathbf{z}}$ is our reduced-order state. Projecting ${\mathbf{z}}$ back to the full space gives approximately $\tilde{\mathbf{x}}$ that allows us to compute the density:
\begin{linenomath*}\begin{equation}
    {\mathbf{x}}(\mathbf{s},t) \approx \mathbf{U}_r(\mathbf{s})\,\mathbf{z}(t) + \bar{\mathbf{x}}(\mathbf{s})
    \label{eq:densityFromROM}
\end{equation}\end{linenomath*}
More details on POD can be found in \citet{mehta2017methodology}.

\subsubsection{Dynamic Mode Decomposition with control}
To enable prediction of the atmospheric density, we develop a linear dynamic model for the reduced-order state $\mathbf{z}$. First, let's consider the full-dimensional case. Since the atmosphere is highly sensitive to the solar activity, we derive a linear system that considers exogenous inputs:
\begin{linenomath*}\begin{equation}
    \mathbf{x}_{k+1} = \mathbf{A} \mathbf{x}_k+ \mathbf{B} \mathbf{u}_k
    \label{eq:linearModel}
\end{equation}\end{linenomath*}
where $\mathbf{u}_k$ is the system input, which in our case are the space weather inputs. The dynamic matrix ${\bf A}$ and input matrix $\mathbf{B}$ can be estimated from output data using the DMDc algorithm. For this, the outputs of the dynamical system \eqref{eq:linearModel} or snapshots, ${\bf x}_k$, are rearranged into time-shifted data matrices. Let ${\bf X}_1$ and ${\bf X}_2$ be the time-shifted matrix of snapshots such that:
\begin{linenomath*}\begin{equation}
{\bf X_1}= \begin{bmatrix}
| & | & & | \\
\mathbf{x}_1 & \mathbf{x}_2 & \cdots & \mathbf{x}_{m-1} \\
| & | & & |
\end{bmatrix}
, \quad
{\bf X_2}= \begin{bmatrix}
| & | & & | \\
\mathbf{x}_2 & \mathbf{x}_3 & \cdots & \mathbf{x}_m \\
| & | & & |
\end{bmatrix}
, \quad
{\bf \Upsilon}= \begin{bmatrix}
| & | & & | \\
{\mathbf{u}}_1 & {\mathbf{u}}_2 & \cdots & {\mathbf{u}}_{m-1} \\
| & | & & |
\end{bmatrix}
\label{eq:X1X2U}
\end{equation}\end{linenomath*}
where $m$ is the number of snapshots and ${\bf \Upsilon}$ contains the corresponding inputs. The data matrices ${\bf X}_1$ and ${\bf X}_2$ are related (${\bf X}_2$ is the time evolution of ${\bf X}_1$) through the model in Eq.~\eqref{eq:linearModel} such that:
\begin{linenomath*}\begin{equation}\label{e:DMD2}
{\bf X}_2 = {\bf A}{\bf X}_1 + \mathbf{B} \mathbf{\Upsilon}
\end{equation}\end{linenomath*}
The goal now is to estimate $\mathbf{A}$ and $\mathbf{B}$. However, because $\mathbf{X}_1$ and $\mathbf{X}_2$ can be extremely large, it is more efficient to perform the DMDc in the reduced-order space:
\begin{linenomath*}\begin{equation}
    \mathbf{Z}_{2} = \mathbf{A}_r \mathbf{Z}_1 + \mathbf{B}_r \mathbf{\Upsilon}
    \label{eq:ROMlinearModel}
\end{equation}\end{linenomath*}
where $\mathbf{Z}_1 = \mathbf{U}_r^T \mathbf{X}_1$ and $\mathbf{Z}_2 = \mathbf{U}_r^T \mathbf{X}_2$ are the reduced-order snapshot matrices, and $\mathbf{A}_r = \mathbf{U}_r^T \mathbf{A} \mathbf{U}_r$ and $\mathbf{B}_r = \mathbf{U}_r^T \mathbf{B}$ are the reduced-order dynamic and input matrices. 
The above equation is modified such that:
\begin{linenomath*}\begin{equation}\label{e:DMD3}
{\bf Z}_2 = {\bf \Xi}\mathbf{\Psi}
\end{equation}\end{linenomath*}
where ${\bf \Xi}$ and $\mathbf{\Psi}$ are the augmented operator and data matrices, respectively:
\begin{linenomath*}\begin{equation}\label{e:DMDc3}
{\bf \Xi} \triangleq \begin{bmatrix} \mathbf{A}_r & \mathbf{B}_r
\end{bmatrix} \quad \text{and} \quad \mathbf{\Psi} \triangleq \begin{bmatrix} {\bf Z}_1 \\ 
\mathbf{\Upsilon}
\end{bmatrix}
\end{equation}\end{linenomath*}
We now estimate the dynamic and input matrices by minimizing $\| {\bf Z}_{2} - {\bf \Xi}\mathbf{\Psi} \|$. The augmented operator matrix is then solved for by computing the pseudoinverse of $\mathbf{\Psi}$:
\begin{linenomath*}\begin{equation}\label{e:DMDc4}
{\bf \Xi}={\bf Z}_2\mathbf{\Psi}^{+}
\end{equation}\end{linenomath*}
where the $+$ subscript indicates the pseudoinverse. In MATLAB, this can be easily computed using the backslash operator: $\mathbf{Z}_2 \mathbf{\Psi}^{+} = \left( (\mathbf{\Psi}^{+})^T \mathbf{Z}_2^T \right)^T = \left( \mathbf{\Psi}^T \backslash \mathbf{Z}_2^T \right)^T $. Calculating $\mathbf{A}_r$ and $\mathbf{B}_r$ without first computing $\mathbf{A}$ and $\mathbf{B}$ is an improvement with respect to previous work \citep{mehta2018quasi} and is also numerically more stable because the number of entries that needs to be determined for matrix $\mathbf{A}_r$ is much smaller than for $\mathbf{A}$.

Finally, the discrete-time matrices are converted to continuous-time matrices to enable continuous-time propagation of the ROM modes for estimation. This can be achieved using the following relation \citep{DeCarlo1989Linear}:
\begin{linenomath*}\begin{equation}\label{e:D2C}
\begin{bmatrix} 
{\bf A}_c & {\bf B}_c\\
{\bf 0} & {\bf 0} 
\end{bmatrix} = \log \left(\begin{bmatrix} 
{\bf A}_d & {\bf B}_d\\
{\bf 0} & {\bf I} 
\end{bmatrix}\right) / {\text T}
\end{equation}\end{linenomath*}
where ${\bf A}_c$ is the dynamic matrix and ${\bf B}_c$ is the input matrix in continuous time, and ${\text T}$ is the sample time, i.e. the snapshot resolution.

\subsubsection{ROM density model development}
\textit{Density training data} In this work, we have developed three different ROM density models using three different atmospheric models to obtain the snapshot matrices, namely the empirical NRLMSISE-00 \citep{PiconeEtAl2002} and Jacchia-Bowman 2008 (JB2008) models \citep{bowman2008new} and the physics-based TIE-GCM model \citep{qian2014ncar}. We first defined a spatial grid $\mathbf{s}$ in local solar time, geographic latitude and altitude and computed the density on this grid for every hour over 12 years (one solar cycle), resulting in over 105,000 snapshots. These snapshots were then used to compute a dynamic ROM model, as described in the previous section, using a reduced order of $r=10$. It should be noted that we computed the variation of the density ${\tilde{\mathbf{x}}}$ by first taking the log base 10 of the density and then subtracting the mean: ${\tilde{\mathbf{x}}} = \log_{10}{\mathbf{x}} - \log_{10}{\bar{\mathbf{x}}}$, where $\mathbf{x}$ and $\bar{\mathbf{x}}$ are the density and mean density on the spatial grid. 

Details on the spatial partitioning and 12 year periods applied for generating the ROM models can be found in Table~\ref{tab:ROMproperties}. Note that the JB2008 ROM model was computed over the years 1999-2010 instead of 1997-2008, because no continuous space weather data was available in the year 1998. 
An improvement with respect to previous work is the development of ROM models that are valid above 450 km altitude, which is the limiting altitude for TIE-GCM. The new ROM models based on NRLMSISE-00 and JB2008 extend up to 700 and 800 km altitude, respectively, see Table~\ref{tab:ROMproperties}. 

\begin{table}[htbp]
\begin{adjustwidth}{-2in}{0in} 
    \caption{ROM characteristics: spatial grid and time period.}
   \label{tab:ROMproperties}
        \centering 
\begin{tabular}{lccccccc}
\hline
Base model & \multicolumn{2}{c}{Local solar time}  & \multicolumn{2}{c}{Latitude} & \multicolumn{2}{c}{Altitude} & Years \\
 & Domain & Partitions & Domain & Partitions & Domain & Partitions \\
\hline
TIEGCM & [0, 24] & 25 & [-87.5, 87.5] & 20 & [100, 450] & 21 & 1997-2008 \\
NRLMSISE-00 & [0, 24] & 24 & [-90, 90] & 20 & [100, 700] & 31 & 1997-2008 \\
JB2008 & [0, 24] & 24 & [-87.5, 87.5] & 20 & [100, 800] & 36 & 1999-2010 \\
\hline
\end{tabular}
\end{adjustwidth}
\end{table}

\textit{Space weather inputs}
The space weather inputs $\mathbf{u}_k$ used in the dynamical model are taken from the inputs required by the original density models, see second column in Table~\ref{tab:ROMSWinputs}. In addition to these default inputs, we added the next-hour values for key space weather indices to improve the DMDc prediction, see third column in Table~\ref{tab:ROMSWinputs}. Finally, a new innovation in this work is the addition of nonlinear space weather terms, such as the square of an index, e.g. $ap^2$, or the multiplication of two different indices, e.g. $ap \cdot F_{10.7}$, see nonlinear inputs in Table~\ref{tab:ROMSWinputs}. The improvement of the DMDc model due to adding nonlinear terms will be discussed in the results section.

\begin{table}[tbp]
\begin{adjustwidth}{-2in}{0in} 
    \caption{ROM space weather inputs: doy=day of year, hr=hour in UTC, GMST=Greenwich mean sidereal time, overbars indicate the 81-day average.}
   \label{tab:ROMSWinputs}
        \centering 
\begin{tabular}{lccc}
\hline
Base model & \multicolumn{1}{c}{Standard inputs}  & \multicolumn{1}{c}{Future inputs} & \multicolumn{1}{c}{Nonlinear inputs} \\
\hline
TIE-GCM$^a$ & $doy, hr, F_{10.7}, \bar{F}_{10.7}, Kp$ & $F_{10.7}, Kp$ & $Kp^2, Kp \cdot F_{10.7}$ \\
NRLMSISE-00$^b$ & $doy, hr, F_{10.7}, \bar{F}_{10.7}, \boldsymbol{ap}^{\dag}$ & $F_{10.7}, \bar{F}_{10.7}, \boldsymbol{ap}^{\dag}$ & $\boldsymbol{ap}^2, ap_{now} \cdot F_{10.7}$ \\
JB2008$^c$ & $doy, hr, F_{10.7}, \bar{F}_{10.7}, S_{10}, \bar{S}_{10}, M_{10}, \bar{M}_{10},$ & $F_{10.7}, S_{10}, M_{10},$ & $DSTDTC^2,$ \\
 & $Y_{10}, \bar{Y}_{10}, DSTDTC, GMST, \alpha_{SUN}, \delta_{SUN} $ & $Y_{10}, DSTDTC$ & $DSTDTC \cdot F_{10.7}$ \\
\hline
\multicolumn{4}{l}{$^{\dag}$ $ap$ indices for the NRLMSISE-00 model consist of 8 ap values for up to 57 hours prior to current time} \\
\multicolumn{4}{l}{$^a$ see \citet{qian2014ncar} }\\
\multicolumn{4}{l}{$^b$ see \citet{PiconeEtAl2002} } \\
\multicolumn{4}{l}{$^c$ see \citet{bowman2008new} }
\end{tabular}
\end{adjustwidth}
\end{table}

\subsection{Density estimation}
The neutral mass density is estimated through the assimilation of two-line element orbital data in the dynamic ROM model. This is achieved by simultaneously estimating the ROM state and the orbit and BC of objects using an unscented Kalman filter.

\subsubsection{Two-line element data}
The US Air Force Space Command publicly distributes the orbital data of thousands of Earth-orbiting objects in the form of two-line element sets. From this TLE data, the state of an object (position and velocity) at any epoch can be extracted using the SGP4/SDP4 models \citep{hoots1980spacetrack,vallado2006revisiting}. Hence, the effect of drag can be observed in TLE orbital data if the drag perturbation is strong enough. 

\begin{figure}[tbp]
\begin{adjustwidth}{-2in}{0in}
     \centering
     \includegraphics[width=1.0\textwidth,trim={0cm 0cm 0cm 0cm},clip] {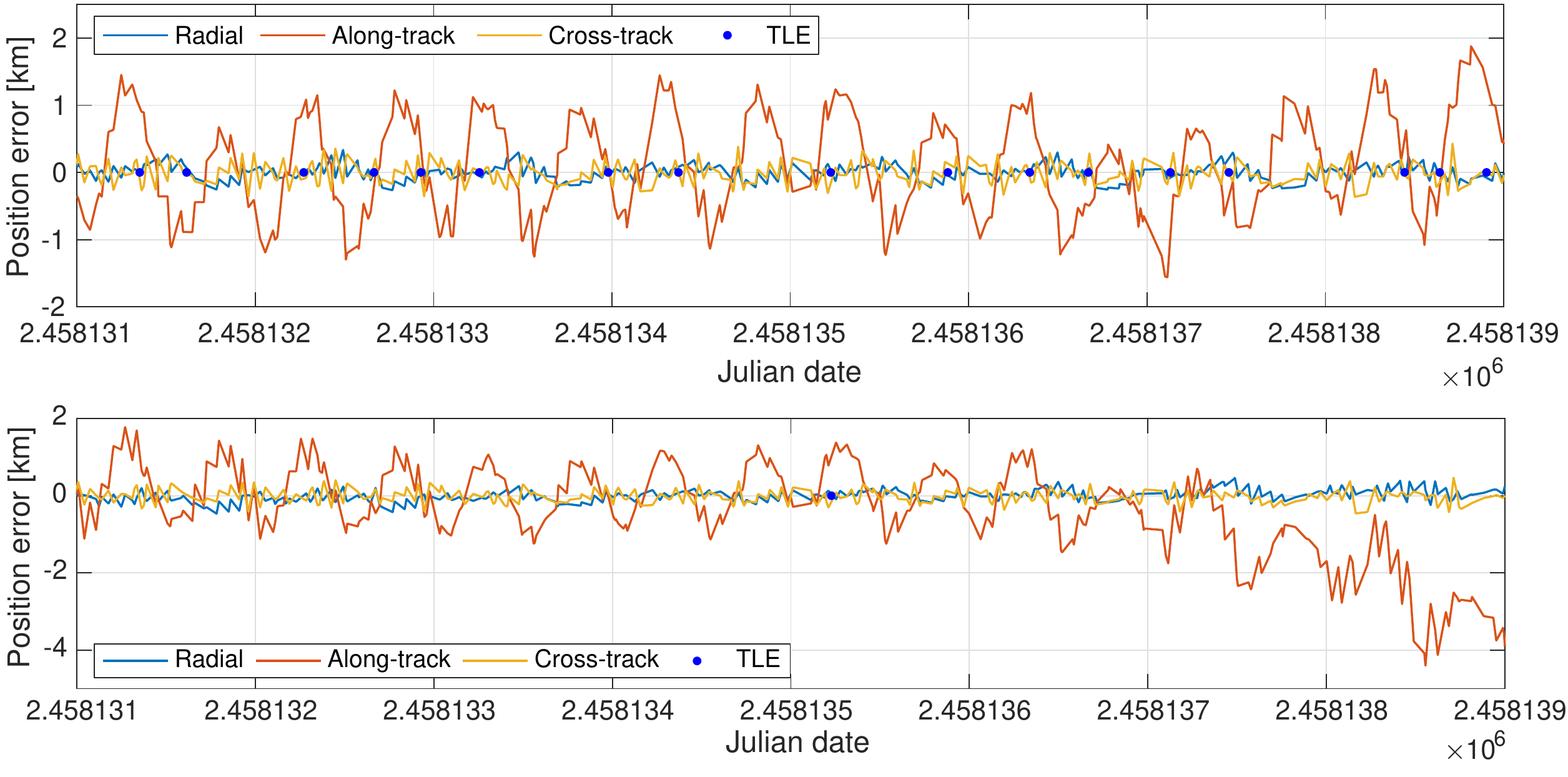}

     \caption{TLE position error with respect to GPS data for a satellite at 494 km altitude using a single TLE (top) or multiple TLEs (bottom) in 8-day window in Jan 2018.}
     \label{fig:TLEvGPS}
\end{adjustwidth}
\end{figure}

A general concern when using TLE data is the accuracy of the orbital data. In the SGP4/SDP4 models only the largest perturbations are included, while higher-order and short-periodic effects, a dynamic atmosphere and orbital maneuvers are not accounted for \citep{vallado2012tlepractice}. As a result, there can be large errors in the orbital data \citep{Kelso2007}. In addition, since 2013, TLEs are fitted to a higher-order orbital solution that includes a future orbit prediction \citep{hejduk2013catalogue}. This generally improve the TLE accuracy at epoch, but may deteriorate the quality if inaccurate future space weather is used for the orbit prediction. 
To gain understanding about errors in TLE data, we compared the position according to TLE data against GPS data for a Planet Labs satellite at 494 km altitude, see Figure~\ref{fig:TLEvGPS}. The position error is largest in the along-track direction and varies with a 12-hour period, which is thought to be due to missing tesseral m-daily terms in the SGP4 theory \citep{herriges1988thesis}. From the figure, it is clear that we can expect significant errors in the orbital data. On the other hand, the errors are expected to be limited when computing the state at epochs prior to the TLE epoch (see top plot in Figure~\ref{fig:TLEvGPS}), which corresponds with the period of tracking data used to generate the TLE.

The objects used for density estimation need to be selected carefully. First of all, the objects preferably have a strong drag signal. In particular, the effect of drag should be strong with respect to other non-conservative force effects, else errors in non-conservative force modelling, such as solar radiation pressure, can result in inaccurate density estimates. Second, it is important to have an accurate estimate of the true BC of the object to minimize errors due to inaccuracies in the BC. Therefore, the variation of the BC over time should preferably be very small or else must be modelled accurately \citep{bowman2002true}. An overview of the objects used in this work is shown in Table~\ref{tab:TLEobjects}. The BC values were taken from \citet{bowman2004method}, \citet{emmert2006thermo} and \citet{lu2017estim}. Finally, any orbit maneuvers or outliers in the TLE data must be detected and excluded from the data before estimation.

\begin{table}[tbp]
\begin{adjustwidth}{-2in}{0in}
    \caption{Objects used for density estimation from 2002 to 2008. The BC values were taken from \citet{bowman2004method}, \citet{emmert2006thermo} and \citet{lu2017estim}.}
   \label{tab:TLEobjects}
        \centering 
   \begin{tabular}{lccccc } 
      \hline 
NORAD & Object  & BC  & Perigee  & Apogee & Inclination \\ 
Catalog ID &  & [m$^2$/kg] &  height [km] &  height [km] &  [deg] \\ 
\hline 
63 & Tiros 2 & 0.01486 & 509 - 460 & 555 - 493 & 48.5 \\
165 & DELTA 1 R/B & 0.05326 & 598 - 532 & 620 - 550 & 47.9 \\
614 & Hitchhiker 1 & 0.01463 & 319 - 312 & 2061 - 1707 & 82.0 \\
2153 & THOR AGENA B & 0.03329 & 501 - 497 & 2636 - 2598 & 79.7 \\
2622 & OV1-9 R/B & 0.02240 & 475 - 471 & 4500 - 4468 & 99.1 \\
4221 & Azur & 0.02201 & 368 - 362 & 1817 - 1586 & 102.7 \\
6073 & Cosmos 482 Debris & 0.00378 & 213 - 206 & 4985 - 3984 & 52.1 \\
7337 & Vektor & 0.01120 & 380 - 376 & 1392 - 1285 & 83.0 \\
8744 & Vektor & 0.01117 & 380 - 372 & 1429 - 1328 & 82.9 \\
12138 & Vektor & 0.01115 & 394 - 390 & 1596 - 1519 & 83.0 \\
12388 & Vektor & 0.01121 & 384 - 379 & 1555 - 1460 & 83.0 \\
14483 & Vektor & 0.01130 & 390 - 385 & 1658 - 1581 & 82.9 \\
20774 & Vektor & 0.01168 & 391 - 387 & 1764 - 1684 & 83.0 \\
23278 & Vektor & 0.01168 & 398 - 394 & 1851 - 1783 & 83.0 \\
26405 & CHAMP & 0.00477$^{\dag}$ & 400 - 314 & 434 - 318 & 87.2 \\
27391 & GRACE 1 & 0.00697 & 480 - 447 & 505 - 470 & 89.0 \\
27392 & GRACE 2 & 0.00693 & 480 - 447 & 506 - 470 & 89.0 \\
      \hline
\multicolumn{6}{l}{$^{\dag}$ Average of values reported by \citet{emmert2006thermo} and \citet{lu2017estim} }
   \end{tabular}
\end{adjustwidth}
\end{table}

\subsubsection{Square-root unscented Kalman filter}
To fuse the model and TLE data, we use the unscented Kalman filter (UKF). The UKF was proposed by \citet{julier1997new} as an extension of the popular Kalman filter \citep{kalman1960new} for application to nonlinear systems. Similar to the extended Kalman filter (EKF), the UKF assumes all variables have Gaussian distributions. However, instead of a first-order linearization of the nonlinear dynamics used by the EKF, the UKF uses an unscented transform (UT) to avoid large errors in the true posterior mean and covariance of the variables. Here, the true posterior mean and covariance are computed to the 3rd order by propagating a carefully selected set of sample points, called sigma points, through the true nonlinear dynamics. The UKF is a popular algorithm that is well documented in literature; therefore, we will only present the algorithm and relevant details. More details about the square-root UKF used in this work can be found in \citet{wan2001unscented}. 

Let us assume a random variable $\textbf{x} \in \mathbb{R}^L$ with mean $\bar{\textbf{x}}$ and covariance $\textbf{P}_\textbf{x}$, that is propagated through a nonlinear function $\pmb f$ such that $\textbf{y} = {\pmb f}(\textbf{x})$. UT uses a set of sigma points to compute the statistics of $\textbf{y}$. This is achieved by generating a matrix $\pmb{\mathcal{X}}$ of 2$L$+1 \textit{sigma vectors} $\mathcal{X}_{i}$ with corresponding weights $W_i$ using the following relationships:
\begin{linenomath*}\begin{equation}\label{e:UT}
\begin{aligned}
\mathcal{X}_{0} &= \bar{\textbf{x}} \\
\mathcal{X}_{i} &= \bar{\textbf{x}} + \sqrt{(L+\lambda)\textbf{P}_\textbf{x})}_{i} \quad i=1,\dots,L\\
\mathcal{X}_{i} &= \bar{\textbf{x}} - \sqrt{(L+\lambda)\textbf{P}_\textbf{x})}_{i-L} \quad i=L+1,\dots,2L\\
W_0^{(m)} &= \lambda/(L+\lambda)\\
W_0^{(c)} &= \lambda/(L+\lambda) + (1-\alpha+\beta)\\
W_i^{(m)} &= W_i^{(c)} = 1/\{2(L+\lambda)\} \quad i=1,\dots,2L\\
\end{aligned}
\end{equation}\end{linenomath*}
where $\lambda = \alpha^2(L+\kappa)-L$ is a scaling parameter, $\alpha$ determines the spread of the sigma points around $\bar{\textbf{x}}$, $\kappa$ is a secondary scaling parameter, and $\beta$ is used to incorporate prior knowledge of the distribution of $\textbf{x}$. Based on the suggested values of the parameters and prior experience, we set the values as $\alpha = 1$, $\beta = 2$, and $\kappa = 3-L$.
The above computed \textit{sigma vectors} are propagated through the nonlinear function:
\begin{linenomath*}\begin{equation}\label{e:UT1}
	\mathcal{Y} = {\pmb f}(\mathcal{X}_i) \quad i =0,\dots,2L
\end{equation}\end{linenomath*}
and the mean and covariance for $\textbf{y}$ are approximated using a weighted sample mean and covariance of the posterior sigma points as follows:
\begin{linenomath*}\begin{equation}\label{e:UT2}
\bar{\textbf{y}} \approx \sum_{i=0}^{2L} W_i^{(m)}\mathcal{Y}_i
\end{equation}\end{linenomath*}
\begin{linenomath*}\begin{equation}\label{e:UT3}
\textbf{P}_{\textbf{y}} \approx \sum_{i=0}^{2L} W_i^{(c)} \{\mathcal{Y}_i - \bar{\textbf{y}}\} \{\mathcal{Y}_i - \bar{\textbf{y}}\}^T
\end{equation}\end{linenomath*}
The UKF extends the UT to recursive estimation through the algorithm given below \citep{wan2001unscented}. The state, dynamics, measurements and noise used to estimate the density with the UKF are described in the following.

\begin{algorithm}
	\caption{Square-Root Unscented Kalman Filter}
	\label{UKFalgorithm}
	
	\textbf{Initialize with:}
	\begin{linenomath*}\begin{equation}\label{e:SQUKF1}
	\hat{\textbf{x}}_0 = \mathbb{E}[\textbf{x}_0] \qquad \textbf{S}_0 = \text{chol} \{\mathbb{E}[(\textbf{x}_0-\hat{\textbf{x}}_0)(\textbf{x}_0-\hat{\textbf{x}}_0)^T]\}
	\end{equation}\end{linenomath*}  
	For $k \in \{1,\dots,\infty\}$,
	\vspace{0.25cm}
	
	\textbf{Sigma point calculation and time update:}
	\begin{linenomath*}\begin{equation}\label{e:SQUKF2}
	\pmb{\mathcal{X}}_k = \left[\hat{\textbf{x}}_k \quad \hat{\textbf{x}}_k \pm \sqrt{(L+\lambda)}\textbf{S}_{k}) \right]
	\end{equation}\end{linenomath*}  
	\begin{linenomath*}\begin{equation}\label{e:SQUKF3}
	\pmb{\mathcal{X}}_{k+1|k} = {\pmb f}\left[\pmb{\mathcal{X}}_k, \textbf{u}_k\right]
	\end{equation}\end{linenomath*}  
	\begin{linenomath*}\begin{equation}\label{e:SQUKF4}
	\hat{\textbf{x}}_{k+1}^- = \sum_{i=0}^{2L}W_i^{(m)}\pmb{\mathcal{X}}_{i,k+1|k}
	\end{equation}\end{linenomath*}
	\begin{linenomath*}\begin{equation}\label{e:SQUKF5}
	\textbf{S}_{k+1}^- = \text{qr} \left\{\left[  \sqrt{W_1^{(c)}}  \left(\pmb{\mathcal{X}}_{1:2L,k+1|k} - \hat{\textbf{x}}_{k+1}^- \right) \quad \sqrt{\textbf{Q}}   \right]\right\} 
	\end{equation}\end{linenomath*}
	\begin{linenomath*}\begin{equation}\label{e:SQUKF6}
	\textbf{S}_{k+1}^- = \text{cholupdate} \left\{ \textbf{S}_{k+1}^- , \pmb{\mathcal{X}}_{0,k+1} - 	\hat{\textbf{x}}_{k+1}^- ,  W_0^{(c)}  \right\} 
	\end{equation}\end{linenomath*}
	\begin{linenomath*}\begin{equation}\label{e:SQUKF7}
	\pmb{\mathcal{Y}}_{k+1|k} = \textbf{H} \left[\pmb{\mathcal{X}}_{k+1|k}\right]
	\end{equation}\end{linenomath*}
	\begin{linenomath*}\begin{equation}\label{e:SQUKF8}
	\hat{\textbf{y}}_{k+1}^- = \sum_{i=0}^{2L}W_i^{(m)}\mathcal{Y}_{i,k+1|k}
	\end{equation}\end{linenomath*}
	\textbf{Measurement update:}
	\begin{linenomath*}\begin{equation}\label{e:SQUKF9}
	\textbf{S}_{\tilde{\textbf{y}}_{k+1}} = \text{qr} \left\{\left[  \sqrt{W_1^{(c)}}  \left(\pmb{\mathcal{Y}}_{1:2L,k+1} - \hat{\textbf{y}}_{k+1} \right) \quad \sqrt{\textbf{R}}   \right]\right\} 
	\end{equation}\end{linenomath*}
	\begin{linenomath*}\begin{equation}\label{e:SQUKF10}
	\textbf{S}_{\tilde{\textbf{y}}_{k+1}} = \text{cholupdate} \left\{ \textbf{S}_{\tilde{\textbf{y}}_{k+1}} , \pmb{\mathcal{Y}}_{0,k+1} - 	\hat{\textbf{y}}_{k+1} ,  W_0^{(c)}  \right\} 
	\end{equation}\end{linenomath*}
	\begin{linenomath*}\begin{equation}\label{e:SQUKF11}
	\textbf{P}_{\textbf{x}_{k+1} \textbf{y}_{k+1}} = \sum_{i=0}^{2L}W_i^{(c)}   \left(\mathcal{X}_{i,k+1|k} - \hat{\textbf{x}}_{k+1}^- \right) \left(\mathcal{Y}_{i,k+1|k} - \hat{\textbf{y}}_{k+1}^- \right)^T
	\end{equation}\end{linenomath*} 
	\begin{linenomath*}\begin{equation}\label{e:SQUKF12}
	\pmb{\mathcal{K}}_{k+1} = \left( \textbf{P}_{\textbf{x}_{k+1} \textbf{y}_{k+1}} /   \textbf{S}_{\tilde{\textbf{y}}_{k+1}}^T \right) / \textbf{S}_{\tilde{\textbf{y}}_{k+1}}
	\end{equation}\end{linenomath*} 
	\begin{linenomath*}\begin{equation}\label{e:SQUKF13}
	\hat{\textbf{x}}_{k+1} = \hat{\textbf{x}}_{k+1}^- + \pmb{\mathcal{K}}_{k+1}\left(\textbf{y}_{k+1} - \hat{\textbf{y}}_{k+1}^- \right)
	\end{equation}\end{linenomath*} 
	\begin{linenomath*}\begin{equation}\label{e:UKF14}
	\textbf{U} = \pmb{\mathcal{K}}_{k+1}\textbf{S}_{\tilde{\textbf{y}}_{k+1}}
	\end{equation}\end{linenomath*} 
	\begin{linenomath*}\begin{equation}\label{e:UKF15}
	\textbf{S}_{k+1} = \text{cholupdate} \left\{ \textbf{S}_{k+1} , \textbf{U} ,  -1  \right\} 
	\end{equation}\end{linenomath*} 
	where \textbf{Q} is the process noise covariance and \textbf{R} is the measurement covariance.
\end{algorithm}

\paragraph{State}
The state $\textbf{x}$ that is estimated in the UKF consists of the osculating orbital states (expressed in modified equinoctial elements) and the BCs of the objects plus the reduced-order density state ${\bf z}$:
\begin{linenomath*}\begin{equation}
\textbf{x}= \begin{bmatrix}
p_1, f_1, g_1, h_1, k_1, L_1,{BC_1}, & ... &, p_n, f_n, g_n, h_n, k_n, L_n,{BC_n}, & {\bf z}^T \\
\end{bmatrix} ^T
\end{equation}\end{linenomath*}
where $n$ is the number of objects and the modified equinoctial elements (MEE) are defined as \citep{Walker1985}:
\begin{linenomath*}\begin{equation}
\begin{array}{lll}
p = a(1-e^2),					&	f = e\cos{(\omega+\Omega)},		&	g = e\sin{(\omega+\Omega)}, \\
h = \tan{(i/2)}\cos{\Omega},~	&	k = \tan{(i/2)}\sin{\Omega},~	&	L = \Omega+\omega+\nu .
\end{array} 
\label{eq:equinoctialelements}
\end{equation}\end{linenomath*}
where $a,e,i,\Omega,\omega$ and $\nu$ are the classical Keplerian orbital elements. The MEE are nonsingular and tend to behave less nonlinear than the Cartesian coordinates (used in previous work) which benefits the Kalman filter estimation. To initialize the state, we use the objects' orbital states according to TLE data and take the BC values from Table~\ref{tab:TLEobjects}. The ROM state is initialized using densities from the JB2008 model. 

\paragraph{Dynamic model}
The dynamic model ${\pmb f}(\textbf{x},t)$ for evolving the state $\textbf{x}$ consists of propagating the orbital states using orbital dynamics and evolving the ROM state using the continuous-time DMDc model:
\begin{linenomath*}\begin{equation}
\dot{\textbf{x}}={\pmb f}(\textbf{x},t) = \begin{bmatrix}
\dot x \\ \dot y \\ \dot z \\ \dot v_x \\ \dot v_y \\ \dot v_z \\ \dot{BC} \\ \dot{\bf z} \\
\end{bmatrix}
=
\begin{bmatrix}
v_x \\ v_y \\ v_z \\
a_{grav,x} + a_{drag,x} \\ 
a_{grav,y} + a_{drag,y} \\
a_{grav,z} + a_{drag,z} \\
0 \\ 
\pmb{\mathbf{A}}_c{\mathbf z} + \pmb{\mathbf{B}}_c{\bf u}
\end{bmatrix}
\end{equation}\end{linenomath*}
where $[x, y, z]$ and $[v_x, v_y, v_z]$ are the inertial position and velocity, respectively, that are used for orbit propagation and $BC = \frac{C_DA}{m}$ is the ballistic coefficient. The ROM state ${\bf z}$ is used to compute the atmospheric density by converting ${\bf z}$ to the full space (see Eq.~\ref{eq:densityFromROM}) and interpolating the density grid. After propagation, the Cartesian state is converted back to MEE, which are used in the UKF.

\paragraph{Orbital dynamics}
\label{sub:orbitDyn}
The orbital dynamics used in this work considers:
\begin{itemize}
  \item Geopotential acceleration computed using the EGM2008 model, up to degree and order 20 for the harmonics;
  \item Atmospheric drag considering a rotating atmosphere for computing the velocity relative to the atmosphere. The atmospheric density is computed using the ROM density model.
\end{itemize}
The orbit propagation is carried out in the inertial J2000 reference frame using Cartesian position and velocity while the geopotential and drag accelerations are computed in the Earth-fixed ITRF93 frame.
NASA's SPICE toolbox is used 
for reference frame and time transformations (ITRF93 and J2000 reference frames and leap-seconds kernel).
Perturbations due to solar radiation pressure, gravitational attraction by the Sun and Moon, and higher-order Earth harmonics are not included, because their effect on the considered orbits during density estimation was found to be negligible compared to other modeling and measurement errors.

\paragraph{Measurements}
The measurements used for estimation are the osculating orbital states extracted from TLE data. 
At one hour intervals the osculating state of each object is computed using the nearest newer TLE by propagating the TLE backward to the measurement epoch using SGP4. These states are then converted to MEE and used as measurements. The 17 objects used in this work for density estimation are shown in Table~\ref{tab:TLEobjects}. Note that for calibrating the ROM-TIEGCM model only 11 objects are used, because 6 of the 17 objects have their perigee above the ROM-TIEGCM maximum altitude of 450 km.

\paragraph{Measurement and process noise}
The measurements noise $\textbf{R}$ was determined empirically. The variance for the measurements of $p$, $f$ and $g$ was scaled by the orbit's eccentricity $e$, because the errors were found to increase with increasing eccentricity:
\begin{linenomath*}\begin{equation}
[R_p, R_f, R_g, R_h, R_k, R_L] = [c_1 \cdot 10^{-8}, c_2 \cdot 10^{-10}, c_2 \cdot 10^{-10}, 10^{-9}, 10^{-9}, 10^{-8}]
\end{equation}\end{linenomath*}
where $c_1 = 1.5\cdot \max(4e,0.0023)$ and $c_2 = 3\cdot \max(e/0.004,1)$.
The process noise variance $\textbf{Q}$ for the state and BC was set to:
\begin{linenomath*}\begin{equation}
[Q_p, Q_f, Q_g, Q_h, Q_k, Q_L, Q_{BC}] = [1.5\cdot 10^{-8}, 2\cdot 10^{-14}, 2\cdot 10^{-14}, 10^{-14}, 10^{-14}, 10^{-12}, 10^{-16}]
\end{equation}\end{linenomath*}
The process noise for the ROM state $\mathbf{Q_z}$ was computed using the 1-hour ROM prediction error on the training data:
\begin{linenomath*}\begin{equation}
    \mathrm{diag}(\mathbf{Q_z}) = \mathrm{diag} \left( \mathrm{Cov}[\mathbf{Z}_{2} - (\mathbf{A}_r \mathbf{Z}_1 + \mathbf{B}_r \mathbf{\Upsilon})] \right)
    \label{eq:ROMprocessNoise}
\end{equation}\end{linenomath*}
As a result of this approach, the Kalman filter will give more confidence to the model prediction with respect to measurements when the ROM prediction on the training data is more accurate. This approach for determining $\mathbf{Q_z}$ was also found to result in good estimates of the uncertainty in the estimated density.
Finally, the initial covariance for the state was set to:
\begin{linenomath*}\begin{equation}
[P_p, P_f, P_g, P_h, P_k, P_L, P_{BC}, P_{z_1}, P_{z_n}] = [R_p, R_f, R_g, R_h, R_k, R_L, (0.005\cdot BC)^2, 20, 5]
\end{equation}\end{linenomath*}
where $P_{z_1}$ refers to the covariance for the first reduced-order mode $z_1$ and $P_{z_n}$ to the covariance for all other modes.
An overview of the reduced-order model density estimation technique is shown below.

\begin{algorithm}
	\caption{ROM density estimation}
	
	\textbf{Reduced-order modeling}
	
	1. Generate density training data $\mathbf{X}$ (hourly density on grid) using physics-based or empirical density model
	
    2. Select reduced order $r$
    
    3. Compute reduced-order model using a SVD of the snapshots $\mathbf{X}$ (Eqs. \ref{eq:variationX}-\ref{eq:orderReduction})
    
    4. Compute DMDc for reduced-order training data (Eqs. \ref{eq:X1X2U}-\ref{e:D2C})

	\textbf{Density estimation}
	
    5. Download TLE data and estimate true BC
    
    6. Select objects with accurate TLEs (check self-consistency) and stable BC (not maneuvering)
    
    7. Generate measurements (orbital states in MEE) every hour from TLEs using SGP4 using nearest newer TLE
    
    8. Estimate ROM modes ${\bf z}$ using unscented Kalman filter (Algorithm~\ref{UKFalgorithm}) by simultaneously estimating the modes and the state and BC of objects
	
\end{algorithm}

\section{Results}
In this section, the performance of ROM model forecasting and density estimation using TLE data is assessed.

\subsection{ROM density prediction}
\begin{figure}
\begin{adjustwidth}{-2in}{0in}
     \centering
     \includegraphics[width=1.1\textwidth,trim={0cm 0cm 0cm 0cm},clip]{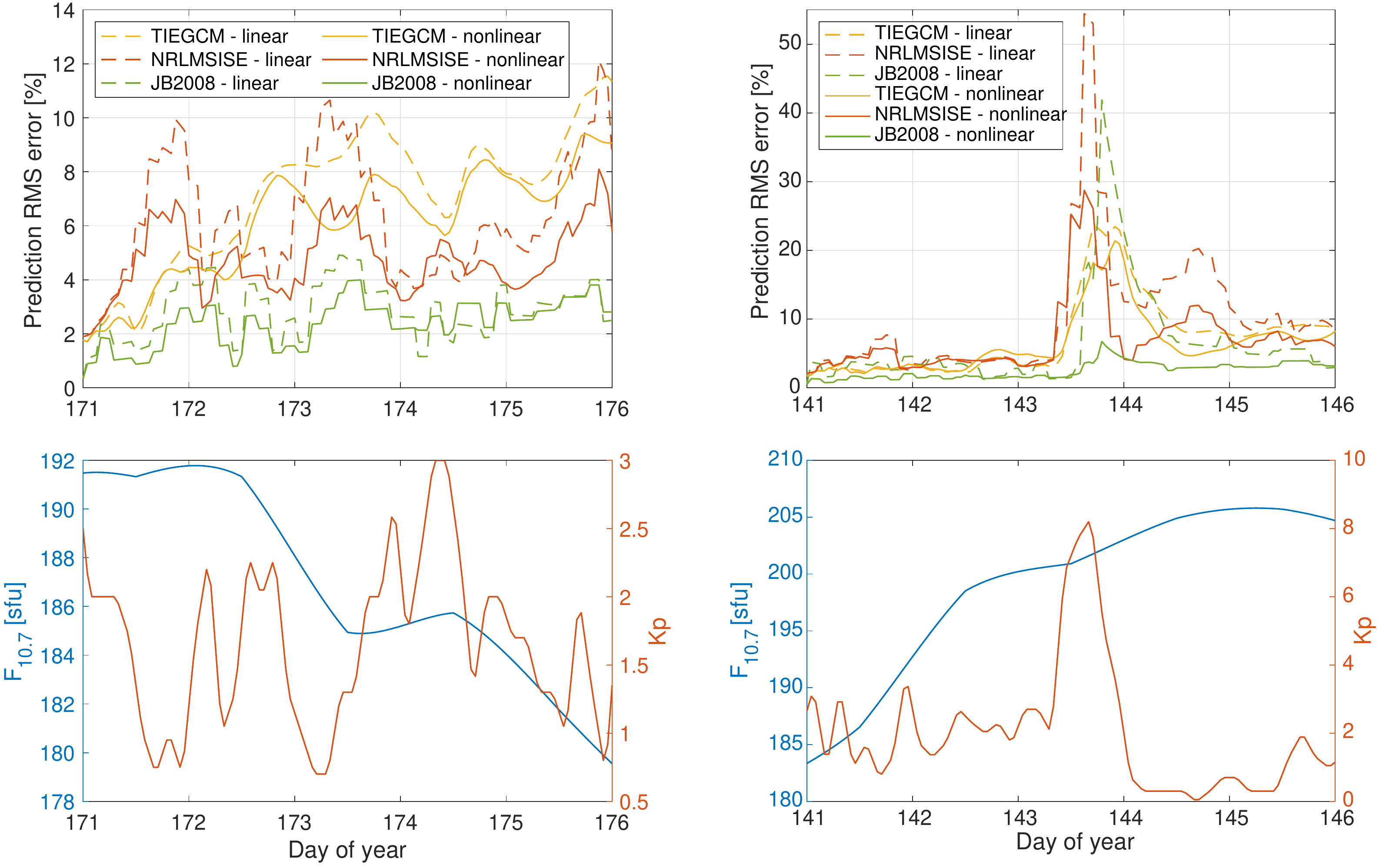}
     \caption{Forecast error in reproducing the 2002 snapshots using different ROM models and using linear or nonlinear space weather inputs during quiet (left) and storm (right) conditions (compare with Figure 5 in \citet{mehta2018quasi}).}
     \label{fig:DMDcForecastError}
\end{adjustwidth}
\end{figure}

The performance of the dynamic ROM models is tested by comparing density forecasts with training data. Using the three different ROM density models the density was predicted for 5 days during quiet space weather conditions and during a geomagnetic storm in 2002. The resulting density forecast errors (the root mean square (RMS) percentage error on the three-dimensional spatial grid) and space weather conditions are shown in Figure~\ref{fig:DMDcForecastError}. The predictions using the ROM model based on JB2008 is most accurate. This good performance can be explained by the superior space weather proxies used by the ROM-JB2008 model. Moreover, the figure and the average 1-hour prediction error in Table~\ref{tab:DMDcPredError} show that the addition of nonlinear space weather terms improves the prediction accuracy for all models. The nonlinear terms especially improve the prediction during a geomagnetic storm. The ROM-JB2008 provides the best predictions with respect to training data; however, this does not necessarily mean it will perform better in estimating true densities.

\begin{table}
\begin{adjustwidth}{-2in}{0in}
    \caption{Average RMS density error (as percentage of true densities) for 1-hr prediction using DMDc models ($r=10$) with linear or nonlinear space weather inputs.}
   \label{tab:DMDcPredError}
        \centering 
   \begin{tabular}{l c c } 
      \hline 
        ROM model & Linear inputs & Nonlinear inputs \\
      \hline 
        TIEGCM & 2.23 & 2.21 \\
        NRLMSISE & 3.47 & 3.38 \\
        JB2008 & 1.32 & 0.90 \\
      \hline
   \end{tabular}
\end{adjustwidth}
\end{table}

\subsection{Simulated TLE test case}
To assess whether accurate density estimation using a ROM model and TLE data is feasible, we first tested the technique using simulated TLE data.
In this scenario, the `true' orbits and densities are first computed by propagating several objects using the ROM-TIEGCM density model. The initial orbital parameters for the true orbits are shown in Table~\ref{tab:simulInitialStates}. Based on these `true' orbits, TLE data is simulated using realistic uncertainties. This simulated TLE data is then used to calibrate the ROM-TIEGCM model. Table~\ref{tab:simulTLEerrors} shows the 1-$\sigma$ errors used for simulated TLE measurements. These errors were established empirically based on TLE data of near-circular orbits around 400 km altitude in the years 2017 and 2018. These errors convert to RMS position and velocity errors of 0.88 km and 1.0 m/s, respectively. The initial guesses for estimating the orbits, BCs and ROM-state also include realistic errors.

\begin{table}
\begin{adjustwidth}{-2in}{0in}
    \caption{Initial orbital parameters for the eight simulated true orbits.}
   \label{tab:simulInitialStates}
        \centering 
   \begin{tabular}{l ccrrrrc } 
      \hline 
 & $a$ [km] & $e$ [-] & $i$ [deg] & $\Omega$ [deg] & $\omega$ [deg] & $M$ [deg] & BC [m$^2$/kg] \\
      \hline 
Object 1 & 6811.031 & 3.011E-3 & 81.208 & 157.262 & 106.464 & 52.070 & 0.0142 \\
Object 2 & 6777.764 & 1.300E-3 & 81.225 & 184.489 & 329.642 & 122.045 & 0.0170 \\
Object 3 & 6810.172 & 1.293E-3 & 81.215 & 187.594 & 112.894 & 78.318 & 0.0168 \\
Object 4 & 6808.532 & 5.124E-4 & 53.014 & 185.496 & 118.205 & 79.004 & 0.0127 \\
Object 5 & 6794.771 & 2.901E-3 & 82.094 & 76.779 & 354.982 & 127.117 & 0.0560 \\
Object 6 & 6785.760 & 4.594E-4 & 97.435 & 67.678 & 86.303 & 88.988 & 0.0220 \\
Object 7 & 6729.365 & 1.619E-3 & 87.251 & 169.664 & 52.108 & 83.135 & 0.0052 \\
Object 8 & 6828.232 & 1.135E-3 & 30.411 & 270.733 & 29.570 & 295.859 & 0.0536 \\
      \hline
   \end{tabular}
\end{adjustwidth}
\end{table}

\begin{table}
\begin{adjustwidth}{-2in}{0in}
    \caption{1-$\sigma$ errors in MEE for simulated TLE measurements.}
   \label{tab:simulTLEerrors}
        \centering 
   \begin{tabular}{l cccccc } 
      \hline 
 & $p$ [km] & $f$ [-] & $g$ [-] & $h$ [-] & $k$ [-] & $L$ [rad] \\
      \hline
1-$\sigma$ error & 0.045 & 2.0E-05 & 2.0E-05 & 2.0E-05 & 2.0E-05 & 1.25E-04 \\
      \hline
   \end{tabular}
\end{adjustwidth}
\end{table}

\begin{figure}
\begin{adjustwidth}{-2in}{0in}
\centering
\includegraphics[width=1.1\textwidth,trim={0cm 0cm 0cm 0cm},clip]{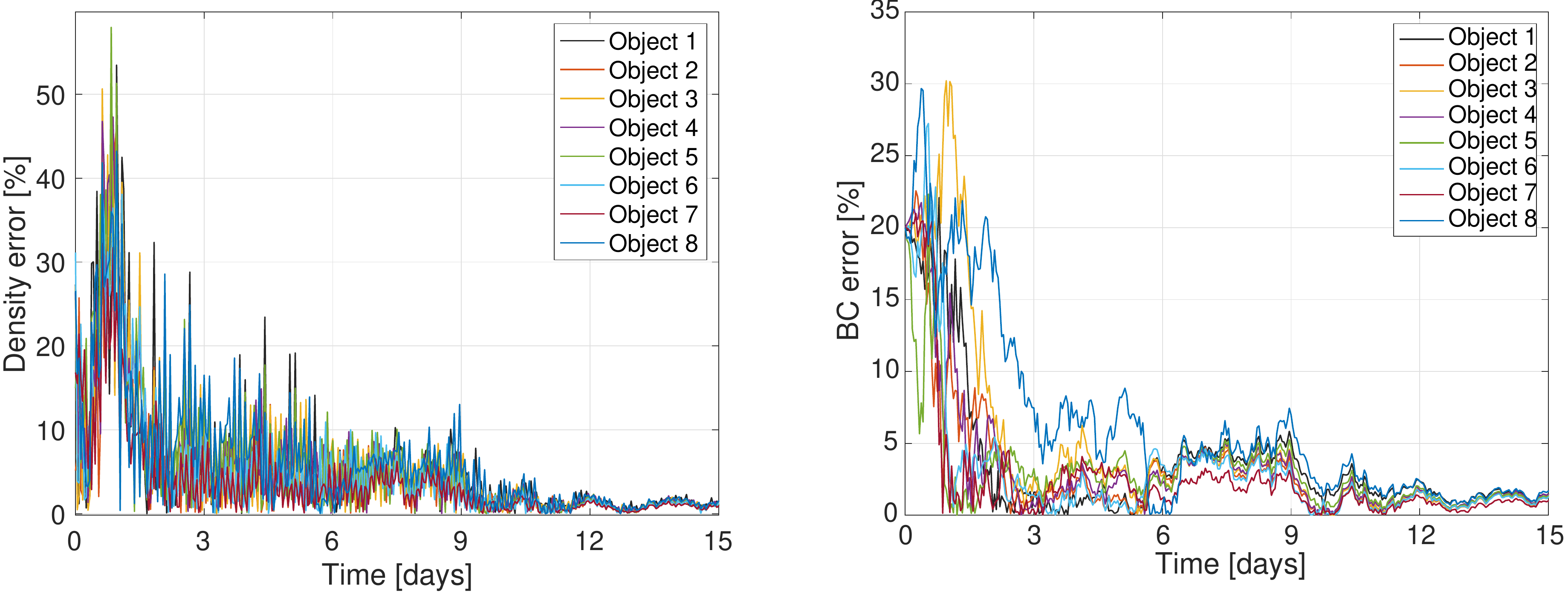}

     \caption{Error in estimated density (left) and BC (right) for each object in simulated TLE test case since 00:00 UTC on day 191 of year 2005.}
     \label{fig:simulDensBCerrors}
\end{adjustwidth}
\end{figure}

\begin{figure}
\begin{adjustwidth}{-2in}{0in}
\centering
\includegraphics[width=1.1\textwidth,trim={3cm 0.3cm 2.7cm 0.5cm},clip]{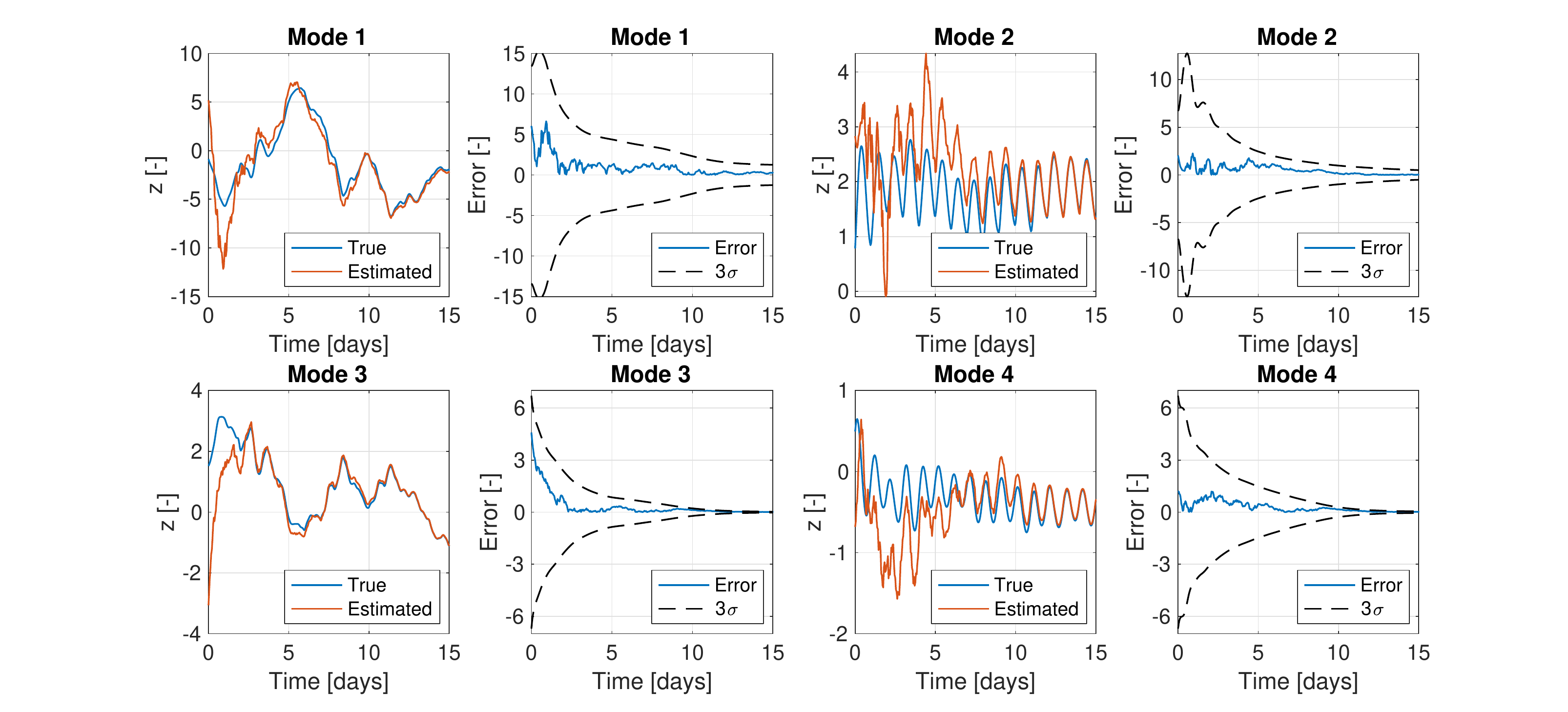}
\caption{True and estimated ROM modes and corresponding error and 3$\sigma$ error bounds for simulated TLE test case since 00:00 UTC on day 191 of year 2005.}
\label{fig:simulROMmodes}
\end{adjustwidth}
\end{figure}

Figure~\ref{fig:simulDensBCerrors} shows the errors in the estimated BCs and densities along the orbits during the estimation period. The errors in density and BC remain below 2\% after 12 days estimation. In Figure~\ref{fig:simulROMmodes} the true and estimated values and errors of the first four ROM modes are shown. All modes converge to their true values, while a small bias in the first mode remains. The convergence of the modes is also correctly displayed by the 3$\sigma$ error bounds. This result demonstrates that accurate density estimation using TLE data is feasible and that both the BC and density are observable. Nevertheless, in reality, less accurate density estimates can be expected, because new TLE measurements are not available every hour and errors in TLE data are not random nor Gaussian.

\subsection{Real TLE}
\begin{figure}[p]
\begin{adjustwidth}{-2in}{0in}
\centering
\includegraphics[width=1.1\textwidth,trim={0cm 0cm 0cm 0cm},clip]{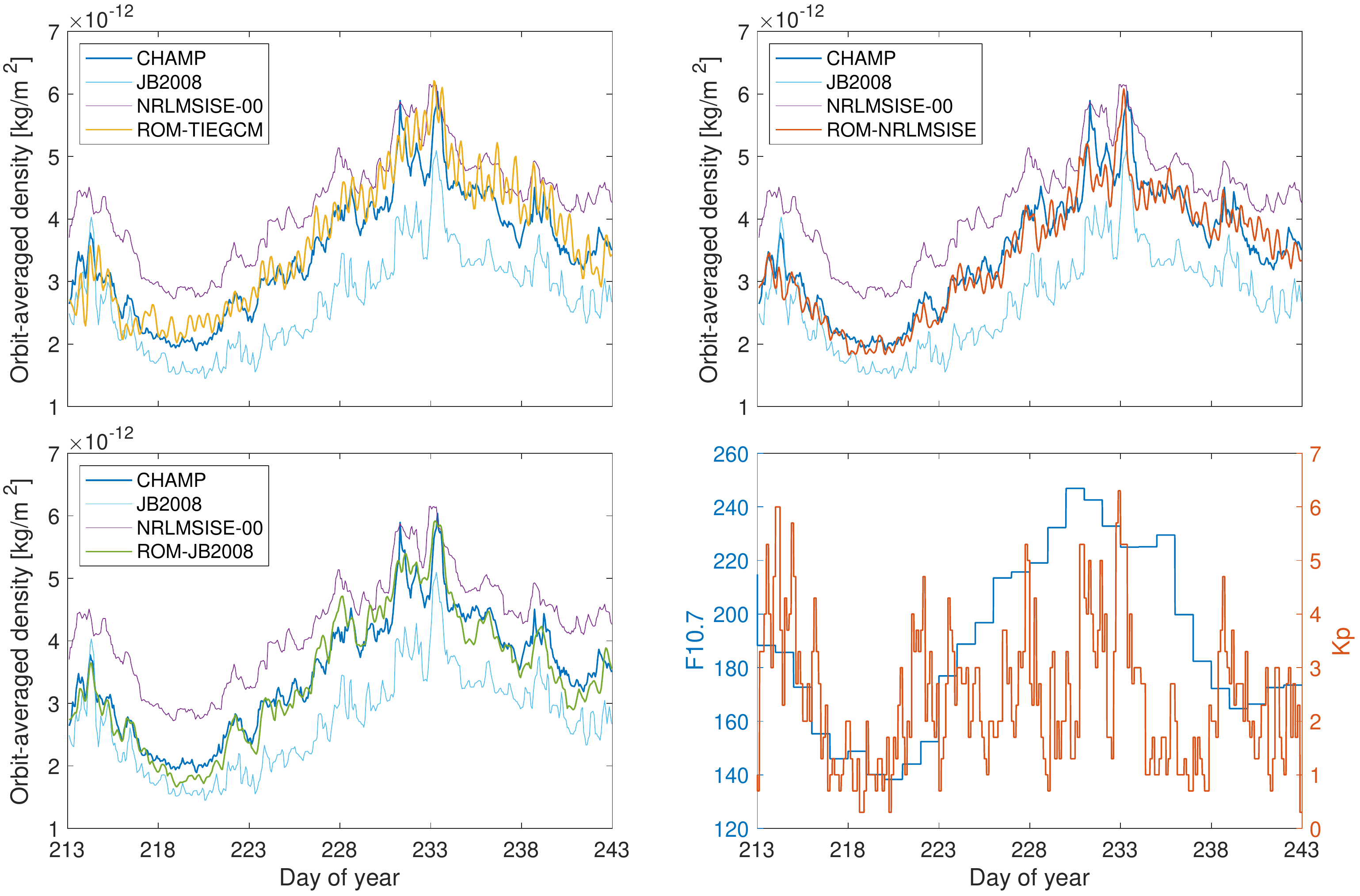}
\caption{Orbit-averaged density along CHAMP orbit according to ROM estimation, CHAMP data, and JB2008 and NRLMSISE-00 models from August 1 to 31, 2002.}
\label{fig:20020801_30d_CHAMP}
\end{adjustwidth}
\end{figure}

In the following, the density is estimated using real TLE data and compared with CHAMP and GRACE accelerometer-derived density data. 
Figure~\ref{fig:20020801_30d_CHAMP} shows the orbit-averaged estimated density along CHAMP's orbit as well as the density according to CHAMP data and the NRLMSISE-00 and JB2008 density models during August 2002 (the first month for which we had both CHAMP and GRACE data). All three ROM models perform very well. Especially, the ROM-NRLMSISE and ROM-JB2008 models are very well able to estimate density variations due to changes in solar activity. The RMS error in daily-averaged CHAMP density is only about 6-9\% for all models, see Table~\ref{tab:densityErrors_Aug0208}. The wiggles in the estimated orbit-averaged density (particularly visible for ROM-TIEGCM and ROM-NRLMSISE) have a period of 12 hours, which suggests that these are due to errors in the TLEs due to missing m-dailies. Similar 12-hour variations can be found in the estimated BCs (not shown here). Further tuning of the measurement and process noise can reduce the amplitude of these variations due to TLE errors.

Overall, the ROM model based on JB2008 performs best with a RMS error in orbit-averaged density of only 7.3\% and 12.1\% along CHAMP's and GRACE-A's orbit, respectively, see Table~\ref{tab:densityErrors_Aug0208}. This shows the high accuracy and temporal resolution that can be achieved by the ROM approach using only TLE data. The error in GRACE-A density is possibly higher because less objects around GRACE's 480 km altitude were used than around CHAMP's altitude of 400 km.
Only 17 objects were used for calibrating the ROM-JB2008 and ROM-NRLMSISE models and only 11 for the ROM-TIEGCM model. This is significantly lower than the 36 and 48 objects used by \citet{doornbos2008use} and \citet{shi2015calibrating}, respectively, but similar to the 16 objects used by \citet{yurasov2005density}.
It should also be noted that the errors presented in this paper are with respect to accelerometer-derived density data and not with respect to TLE-derived density data as in some other works \citep{doornbos2008use,shi2015calibrating}.

\begin{table}[p]
\begin{adjustwidth}{-2in}{0in} 
    \caption{Accuracy of estimated and modelled densities along CHAMP's and GRACE-A's orbit during August 2002. The numbers show RMS difference between orbit-averaged modelled/estimated densities and true densities as percentage of true densities.}
  \label{tab:densityErrors_Aug0208}
        \centering 
  \begin{tabular}{llccccc} 
      \hline 
Satellite & Orbit/Daily & ROM & ROM & ROM & JB2008 & NRLMSISE-00 \\
 & averaged & TIEGCM & NRLMSISE & JB2008 &  &  \\ \hline
\multirow{2}{*}{CHAMP} & Orbit & 11.2 & 8.3 & 7.3 & 23.7 & 26.7 \\
 & Daily & 8.8 & 5.5 & 5.8 & 23.2 & 26.0 \\ \cline{2-7}
\multirow{2}{*}{GRACE-A} & Orbit & -- & 13.3 & 12.1 & 23.4 & 43.1 \\
 & Daily & -- & 10.6 & 9.8 & 22.6 & 42.2 \\ \hline
  \end{tabular}
  \end{adjustwidth}
\end{table}

\begin{figure}[p]
\begin{adjustwidth}{-2in}{0in}
\centering
\includegraphics[width=1.1\textwidth,trim={3cm 0cm 2cm 0.3cm},clip]{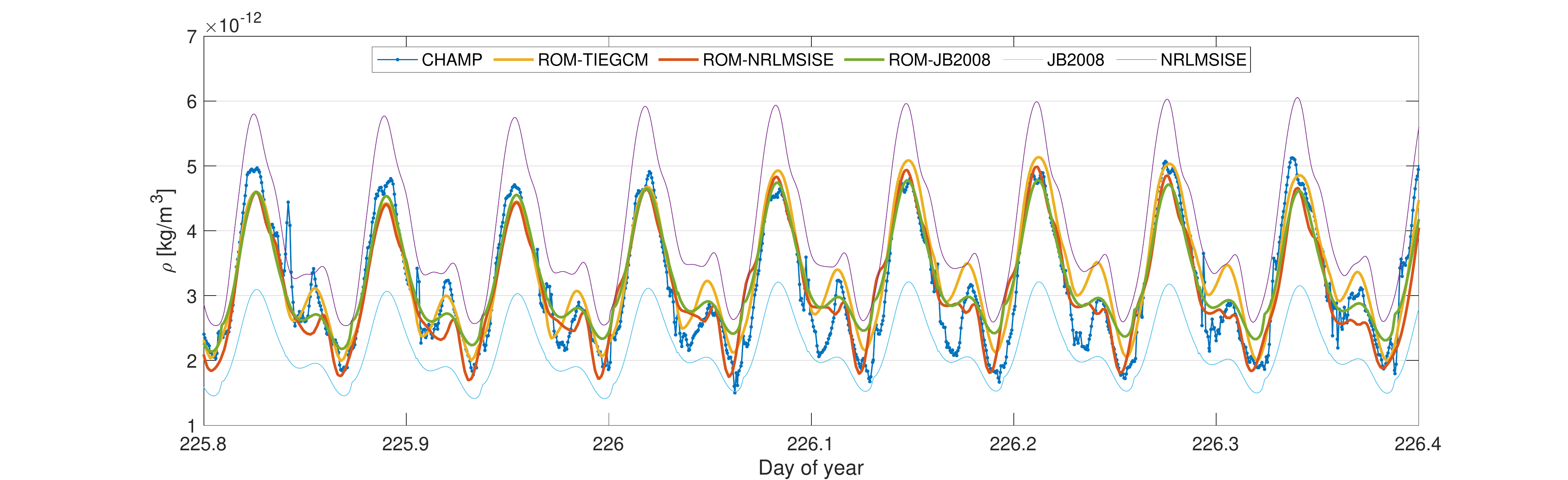}
\caption{Density along CHAMP orbit according to ROM estimation, CHAMP data, and JB2008 and NRLMSISE-00 models (day 226 of year 2002 is August 14, 2002).}
\label{fig:DensityCHAMPzoom}
\end{adjustwidth}
\end{figure}

\begin{figure}
\begin{adjustwidth}{-2in}{0in}
\centering
\includegraphics[width=1.1\textwidth,trim={1cm 0.5cm 0cm 0cm},clip]{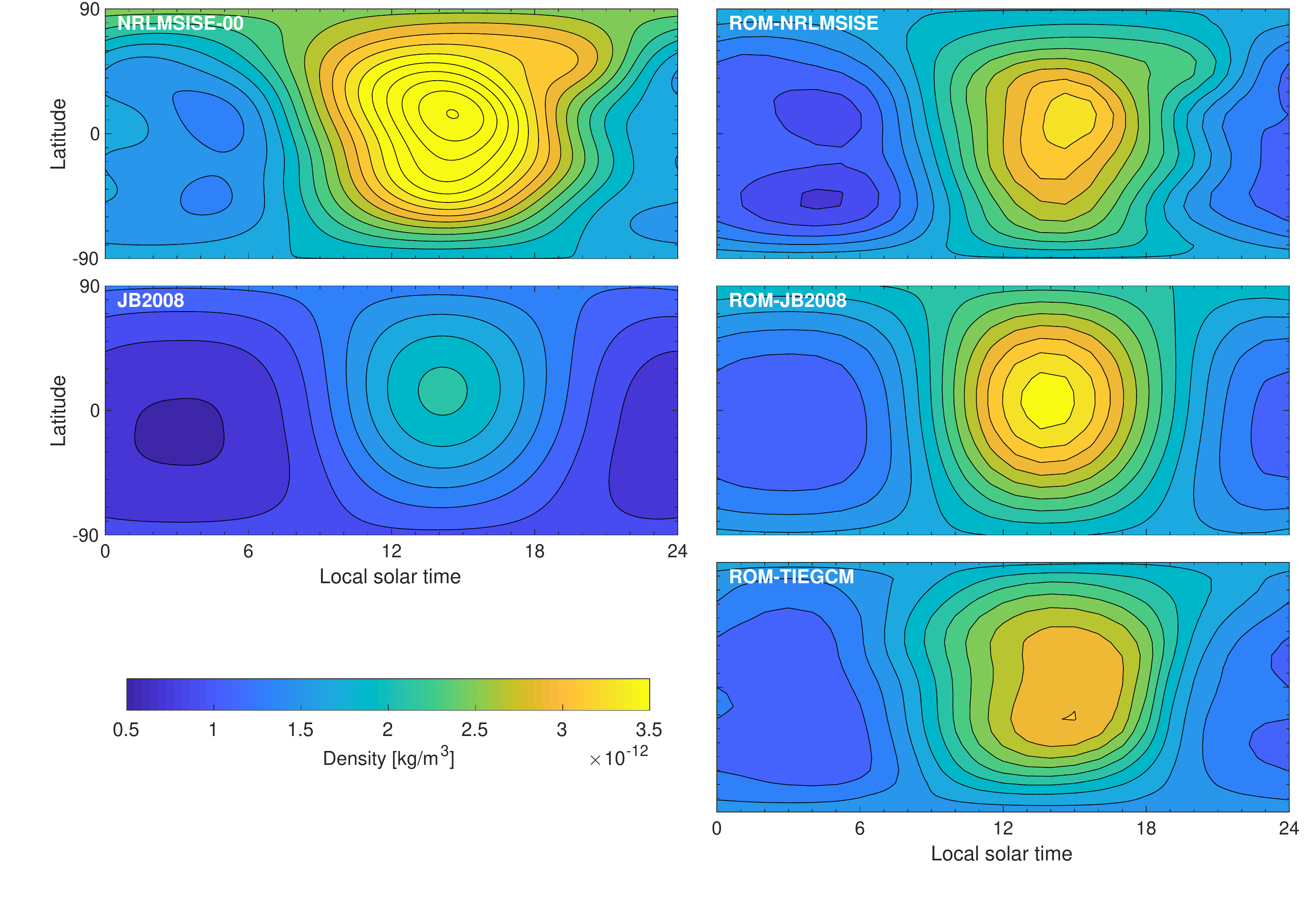}
\caption{Maps of modeled density at 450 km latitude on August 8, 2002 at 0:00:00 UTC.}
\label{fig:densityMaps}
\end{adjustwidth}
\end{figure}

A close-up of the density along CHAMP's orbit on August 14, 2002 is shown in Figure~\ref{fig:DensityCHAMPzoom}. In this time window, the ROM-estimated densities are very close to the true density (both the mean and variation are estimated well). It is probably not feasible to exactly match the true density using TLE data due to the highly-dynamic character of the atmosphere and low-temporal resolution of TLE data. Besides the good match, one can see differences between the different ROM models in the density variation over one orbit. These differences stem from the underlying base models which result in different spatial modes for each model. The way that the ROM models mimic their base model can also be seen in Figure~\ref{fig:densityMaps} that shows maps of the modelled and estimated density on August 8, 2002 at 450 km altitude. For example, the simple density distribution in the JB2008 model is also visible in the ROM-JB2008 density, whereas the NRLMSISE-00 and TIEGCM based ROM models show more complex density distributions. On the other hand, independent of the base model, the ROM-estimated densities have a similar magnitude, which indicates successful calibration.

\begin{figure}
\begin{adjustwidth}{-2in}{0in}
     \centering
     \includegraphics[width=1.1\textwidth,trim={0cm 0cm 0cm 0cm},clip]{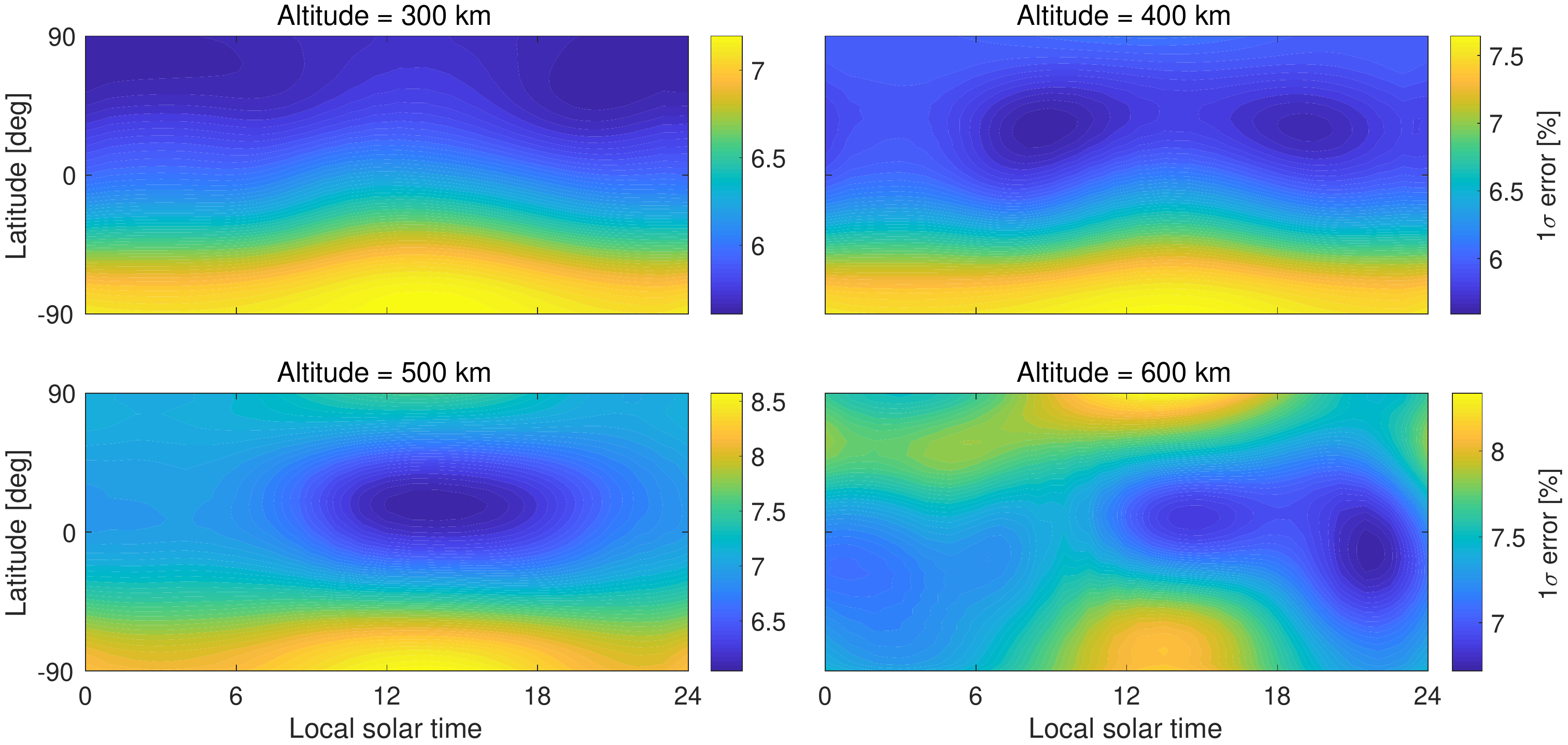}
     \caption{1-$\sigma$ uncertainty in density estimated using ROM-JB2008 at 300 to 600 km altitude after 30 days estimation on August 31, 2002.}
     \label{fig:densityCov}
\end{adjustwidth}
\end{figure}

\begin{figure}
\begin{adjustwidth}{-2in}{0in}
     \centering
     \includegraphics[width=0.66\textwidth,trim={0cm 0cm 0cm 0cm},clip]{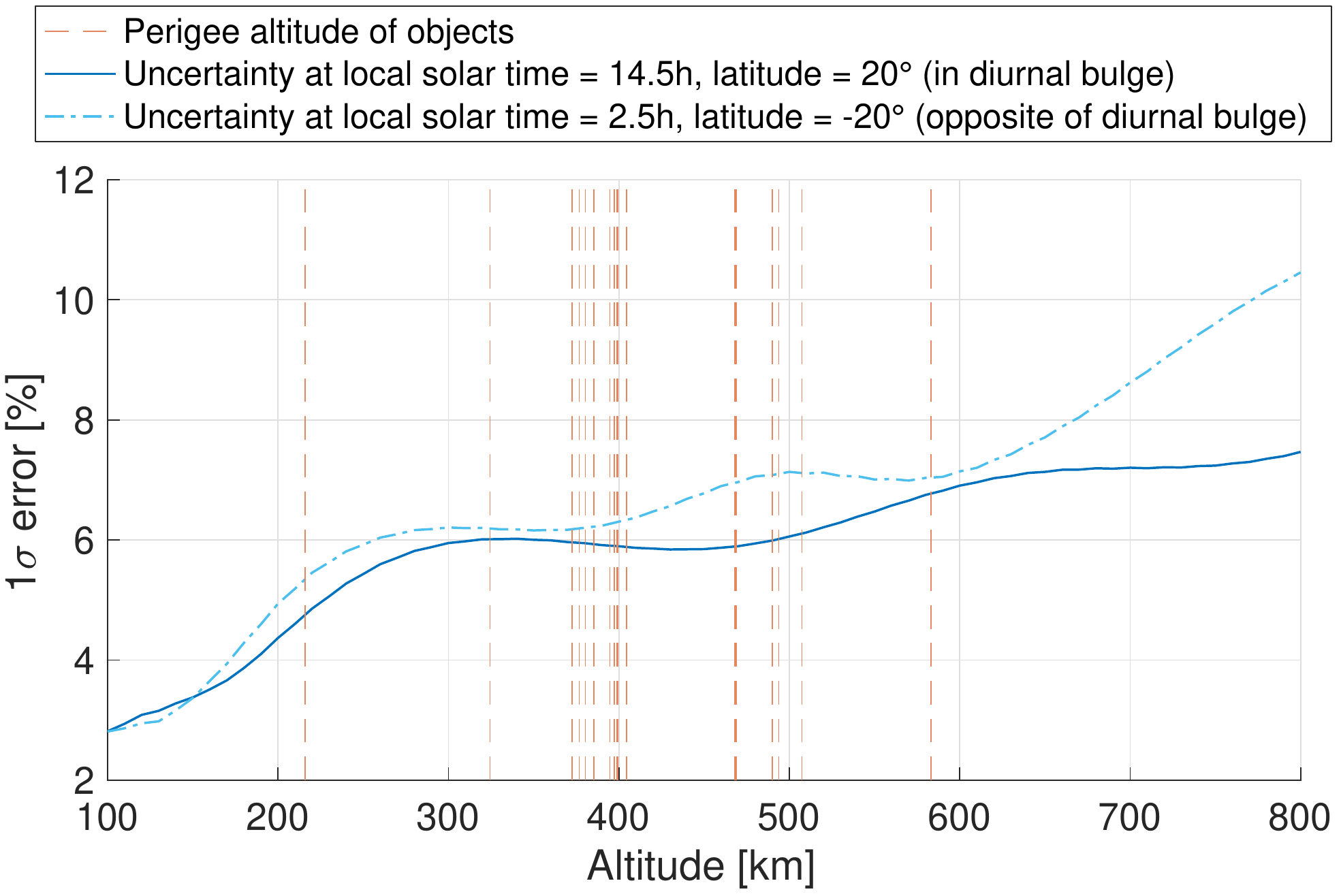}
     \caption{1-$\sigma$ uncertainty in density estimated using ROM-JB2008 at two locations for varying altitude after 30 days estimation on August 31, 2002.}
     \label{fig:densityCov2}
\end{adjustwidth}
\end{figure}

\subsubsection{Uncertainty}
Figures \ref{fig:densityCov} and \ref{fig:densityCov2} show the uncertainty in the estimated density. The plots show that the error in the estimated density is smaller for lower altitude and inside the diurnal bulge. This indicates that the density estimation is more accurate when the drag signal is stronger. In addition, Figure~\ref{fig:densityCov2} shows that the uncertainty grows little at altitudes where measurements are available.
The 1-$\sigma$ error varies between 3 and 11\%, while \citet{mehta2018new} found a 1-$\sigma$ error of 5\% along CHAMP's orbit and up to 25\% error at higher altitudes after assimilating CHAMP density data in a TIEGCM-based ROM model. This indicates that the use of data from multiple objects improves the global density estimates.

\subsubsection{Full years 2003 and 2007}
\begin{figure}
\begin{adjustwidth}{-2in}{0in}
\centering
\includegraphics[width=1.3\textwidth,trim={0cm 0cm 0cm 0cm},clip]{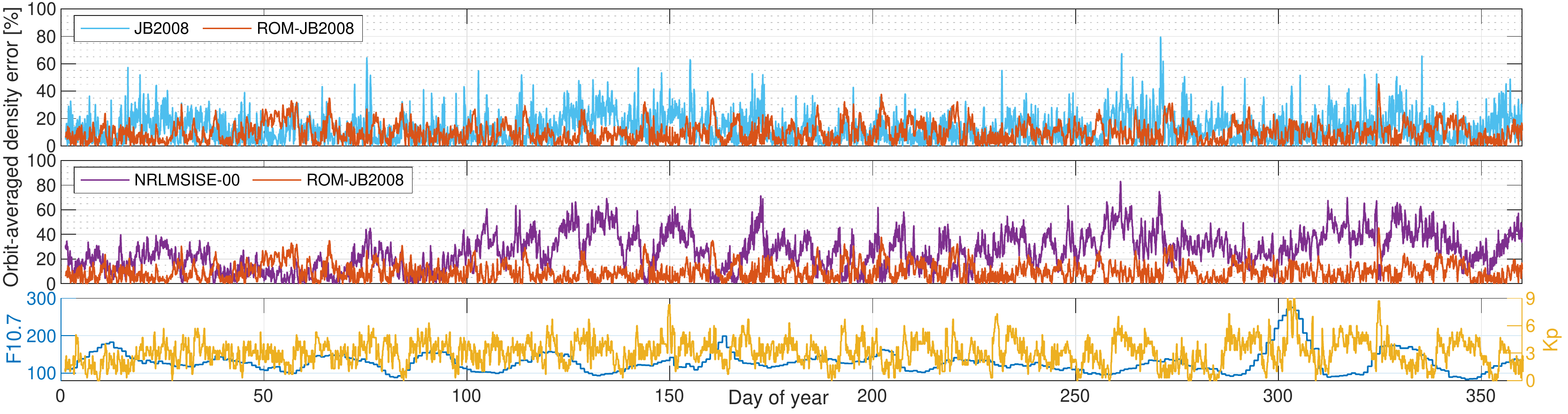}
\caption{Error in orbit-averaged density with respect to CHAMP data for ROM-JB2008 estimated density, and JB2008 and NRLMSISE-00 models for the year 2007.}
\label{fig:2003densities}
\end{adjustwidth}
\end{figure}

The neutral mass density was estimated using the ROM-JB2008 model over the entire years 2003 (high solar activity) and 2007 (low solar activity). The difference in the estimated and true densities along CHAMP's and GRACE-A's orbits are shown in Table~\ref{tab:2003densityErrors} and Figure~\ref{fig:2003densities}. Both the ROM estimation and JB2008 model perform very well in 2003, whereas the ROM estimates are much more accurate than the JB2008 and NRLMSISE-00 models in 2007. The accuracies reported here can be further improved by using more accurate orbital data and by improving the spatial and temporal coverage of the measurements.

\subsubsection{Geomagnetic storm}
\begin{figure}
\begin{adjustwidth}{-2in}{0in}
     \centering
     \includegraphics[width=1.1\textwidth,trim={0cm 0cm 0cm 0cm},clip] {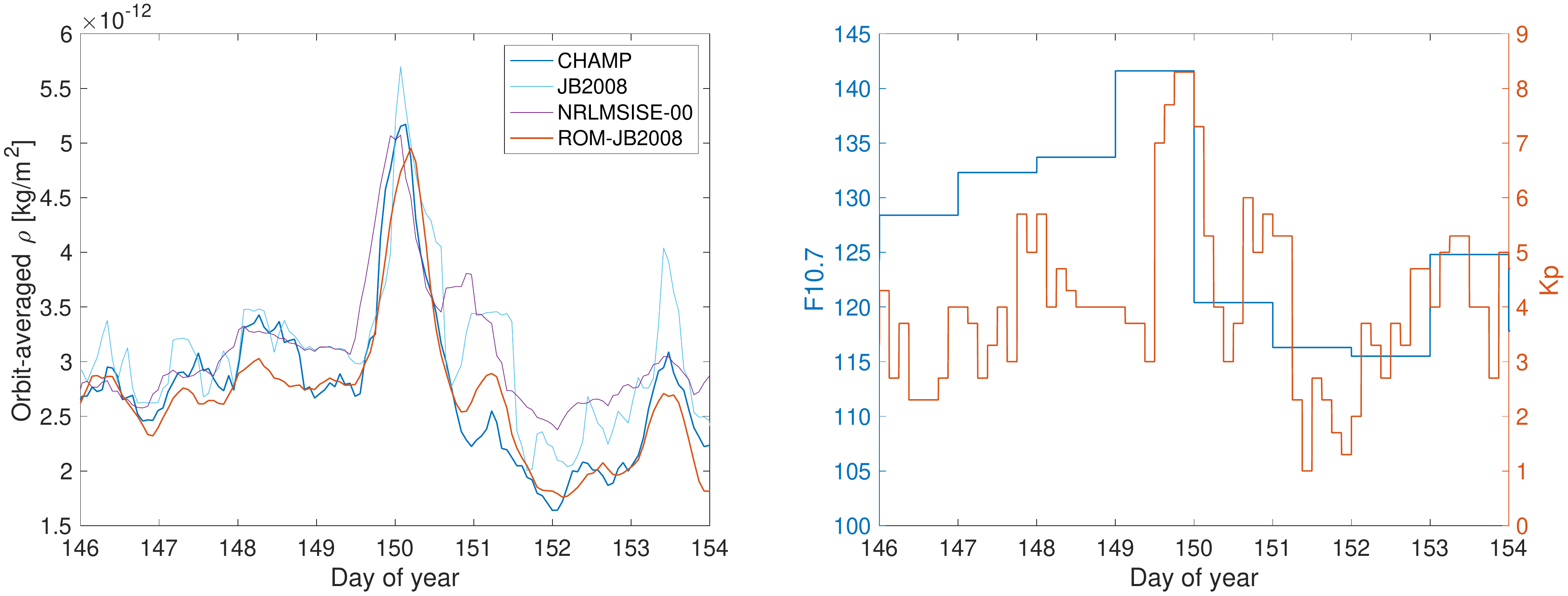}

     \caption{Orbit-averaged density along CHAMP's orbit according to ROM estimation, CHAMP data, and JB2008 and NRLMSISE-00 models during major geomagnetic storm on May 30, 2003 (Kp up to 8.3).}
     \label{fig:ROMdensityStormCHAMP}
\end{adjustwidth}
\end{figure}

Figure~\ref{fig:ROMdensityStormCHAMP} shows the density along CHAMP's orbit estimated using the ROM-JB2008 model during a major geomagnetic storm in 2003. Both the ROM and empirical models provide good density estimates during the storm. However, the empirical models overestimate the density after storm on day 151 (which results in large relative density errors), whereas the ROM estimates are much more accurate. This example shows that the linear ROM model is able to deal with space weather events even though these events are strongly nonlinear. 

\begin{table}
\begin{adjustwidth}{-2in}{0in}
	\fontsize{10}{10}\selectfont
    \caption{Accuracy of estimated and modelled densities along CHAMP's and GRACE-A's orbit over the year 2003 and 2007. The numbers show RMS difference between orbit-averaged modelled/estimated densities and true densities as percentage of true densities.}
  \label{tab:2003densityErrors}
        \centering 
  \begin{tabular}{llccc}
      \hline 
Year & Satellite & ROM-JB2008 & JB2008 & NRLMSISE-00 \\
      \hline
\multirow{2}{*}{2003} & CHAMP   & 12.6	& 12.7	& 20.2 \\
 & GRACE-A & 17.9	& 20.9	& 35.3 \\ \cline{2-5}
\multirow{2}{*}{2007} & CHAMP   & 11.9	& 16.5	& 30.5 \\
 & GRACE-A & 22.4	& 32.7	& 52.5 \\
      \hline
  \end{tabular}
\end{adjustwidth}
\end{table}

\subsection{Including density data}
\begin{figure}
\begin{adjustwidth}{-2in}{0in}
     \centering
     \includegraphics[width=0.55\textwidth,trim={0cm 0cm 0cm 0cm},clip] {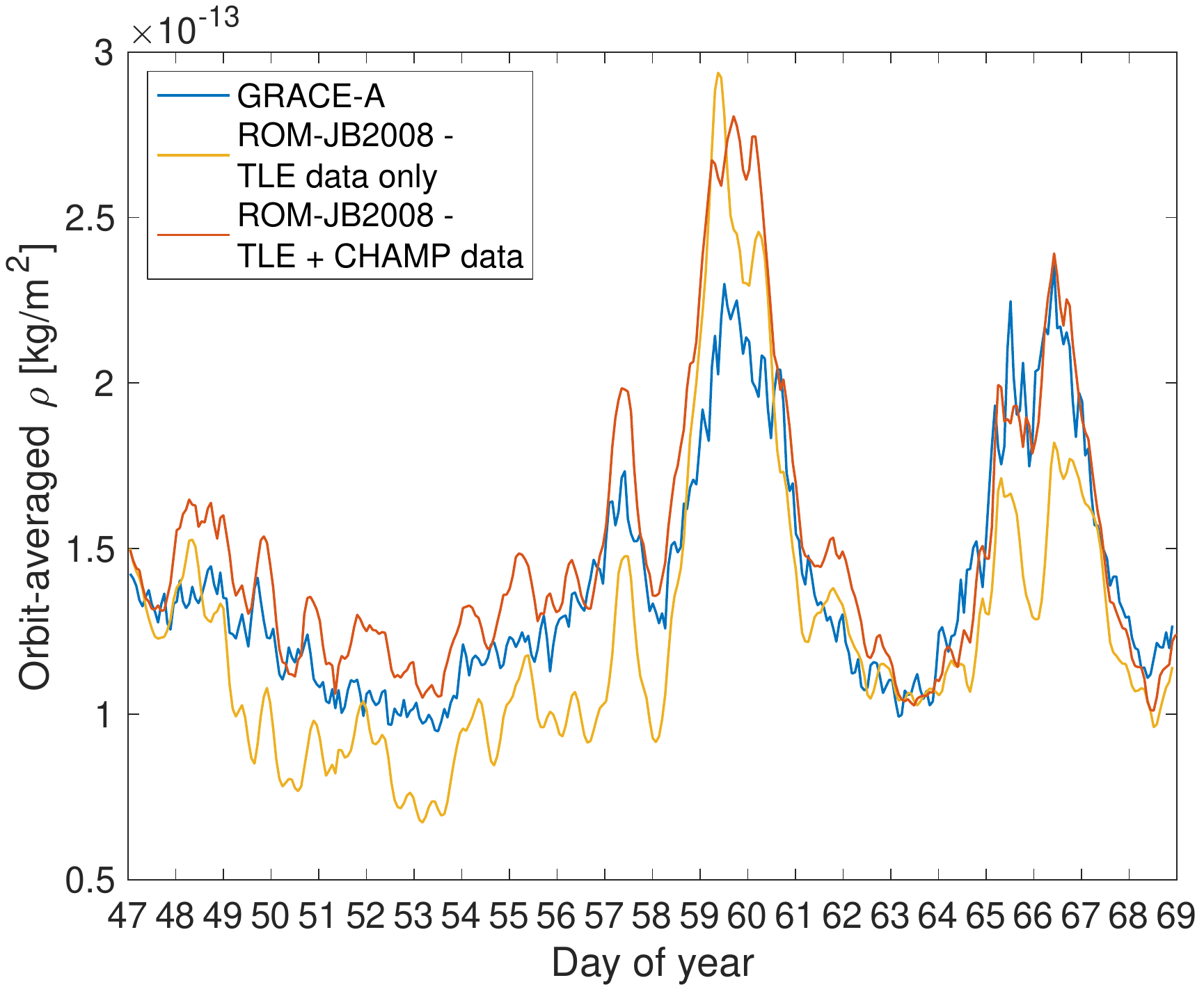}

     \caption{Orbit-averaged density along GRACE-A's orbit estimated using only TLE data or using TLE data and CHAMP accelerometer-derived density data (Feb 16 to Mar 10, 2007).}
     \label{fig:includingCHAMPdata_GRACE}
\end{adjustwidth}
\end{figure}

For some periods of time, one may have access to highly-accurate density data, such as CHAMP, GRACE or GOCE accelerometer-derived densities. This data can be included in the data assimilation to improve the global density estimates. Figure \ref{fig:includingCHAMPdata_GRACE} shows the estimated orbit-averaged density along GRACE-A's orbit after assimilating CHAMP density data together with TLE data (here CHAMP was at 360 km and GRACE-A at 480 km altitude). This period coincides with the period of reduced estimation accuracy shown in Figure~\ref{fig:2003densities} around day 50. One density measurement is included each hour at the same time as the TLE orbit measurements. The inclusion of the CHAMP densities significantly improves the GRACE density estimates; the error in daily-averaged density reduced from 16.4\%
to 11.6\%. Therefore, global density estimates can be improved by including accurate density data. This is especially useful in case the drag signal is weak and consequently the density is difficult to observe from orbital data such as during periods of low solar activity.

\subsection{Density forecast}
\begin{figure}[t]
\begin{adjustwidth}{-2in}{0in}
     \centering
     \includegraphics[width=1.1\textwidth,trim={0cm 0cm 0cm 0cm},clip] {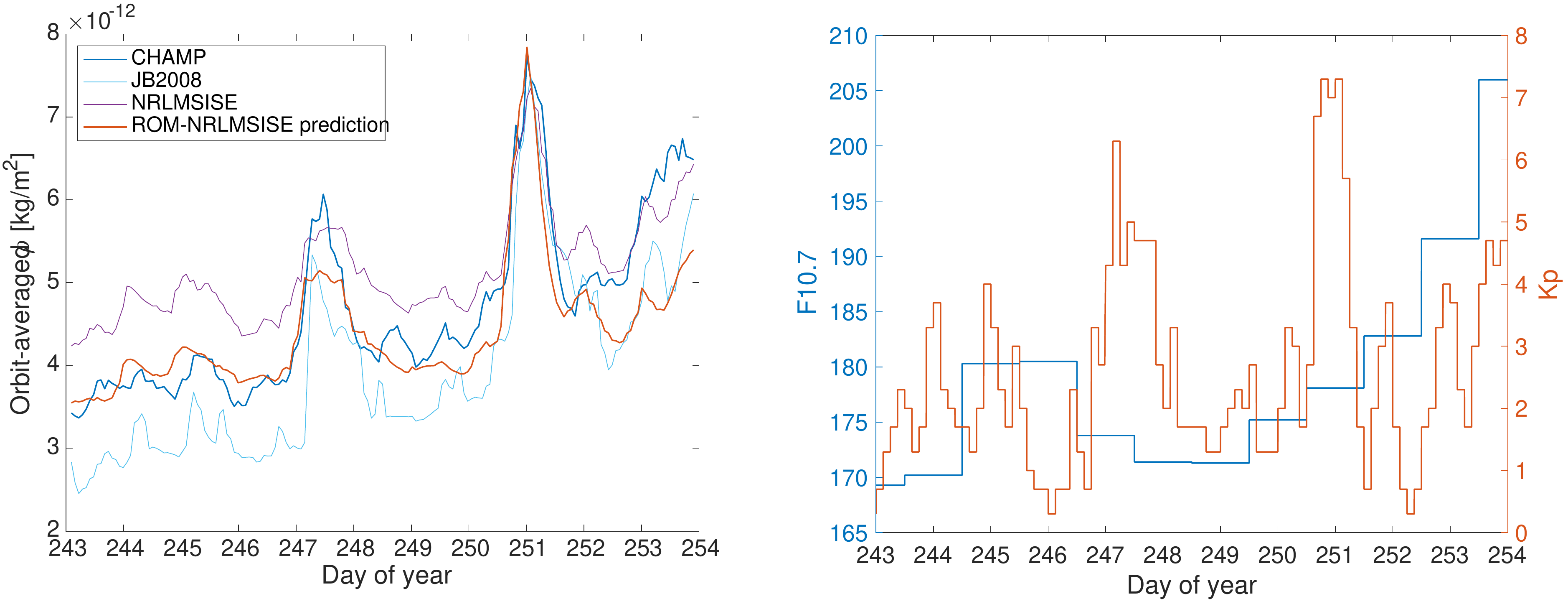} \label{fig:ROMpredictionCHAMPdens}
     \caption{Orbit-averaged density along CHAMP orbit according to ROM prediction, CHAMP data, and JB2008 and NRLMSISE-00 models and $F_{10.7}$ and $Kp$ indices (Aug 31 to Sep 10, 2002).}
     \label{fig:ROMpredictionCHAMP}
     \end{adjustwidth}
\end{figure}

The density along CHAMP's orbit was predicted for 11 days using the ROM-NRLMSISE model, see Figure~\ref{fig:ROMpredictionCHAMP}. The initial ROM state used for the prediction was obtained after 30 days calibration in August 2002, see Figure~\ref{fig:20020801_30d_CHAMP}. The ROM does very well in predicting the density, even during two geomagnetic storms. This example shows that ROM models are able to accurately forecast the future density if the initial state of the thermosphere and the future space weather are accurately known. In future work, prediction of the future space weather will also be considered.

\section{Conclusions}
In this paper we have presented the development of a dynamic ROM model for the thermosphere and estimated its state using TLE data. Three different models based on TIE-GCM, NRLMSISE-00 and JB2008 were generated with an upper altitude up to 800 km. The prediction performance of the models was improved by including nonlinear space weather terms as inputs for the DMDc. In addition, the estimation using an UKF was improved by expressing the measurements in modified equinoctial elements and by calculating the process noise for the ROM model based on training data performance.
Densities were estimated using TLE data and compared with CHAMP and GRACE accelerometer-derived density data. The results showed that the dynamic model enables accurate density estimation using the TLEs of only a small number of satellites. Improved global density estimates can be obtained by including accurate density measurements. Finally, the ROM was shown to be able to provide accurate density forecasts.

Future work will focus on further improving the ROM dynamic models to deal with nonlinearities by improving the choice of space weather inputs and by applying Koopman operator theory.
In addition, we can estimate the global thermospheric neutral density using historic TLE data from the 1960's up to present time to generate a density database.

\section*{Acknowledgments}
The authors wish to acknowledge support of this work by the Air Force's Office of Scientific Research under Contract Number FA9550-18-1-0149 issued by Erik Blasch. The authors would also like to thank Planet Labs, Inc. and Vivek Vittaldev for providing the GPS ephemerides data used in this work. The NRLMSISE-00 and JB2008 models used in this work can be found on \url{https://www.brodo.de/space/nrlmsise} and \url{http://sol.spacenvironment.net/jb2008/code.html}, respectively. The space weather proxies were obtained from \url{http://celestrak.com/SpaceData/} and \url{http://sol.spacenvironment.net/jb2008/indices.html}. The CHAMP and GRACE densities used in this work were derived by \citet{sutton2008effects} and can be found, e.g., on \url{http://tinyurl.com/densitysets}.


\bibliography{references}

\end{document}